\documentclass[fleqn,usenatbib]{mnras}

\usepackage{mathptmx}
\usepackage{txfonts}
\usepackage[T1]{fontenc}
\usepackage{ae,aecompl}
\usepackage[normalem]{ulem}
\usepackage{multirow}

\DeclareRobustCommand{\VAN}[3]{#2}
\let\VANthebibliography\thebibliography
\def\thebibliography{\DeclareRobustCommand{\VAN}[3]{##3}\VANthebibliography}

\usepackage{graphicx}	
\usepackage{subcaption}
\usepackage{xcolor}

\definecolor{OliveGreen}{rgb}{0,0.6,0}

\setcitestyle{sort, super, square} 

\title[The Most Common Habitable Planets III]{The Most Common Habitable Planets III – Modeling Temperature Forcing and Surface Conditions on Rocky Exoplanets and Exomoons}
\author[Beatriz B. Siffert et al.]{Beatriz B. Siffert$^{1}$\thanks{E-mail: beatriz@caxias.ufrj.br},
Raquel G. Gonçalves Farias$^{1,2}$\thanks{E-mail: raquelfarias@usp.com},
Matias Garcia$^{3}$,
Luiz Felipe Melo de Menezes$^{3}$, 
\newauthor{Gustavo F. Porto de Mello$^{4}$, Marcelo Borges Fernandes$^{3}$ and
Rafael Pinotti$^{5}$}
\\
$^{1}$Campus Duque de Caxias, Universidade Federal do Rio de Janeiro, Rod. Washington Luiz 19593, 25240-005, Duque de Caxias, RJ, Brazil\\
$^{2}$Instituto de Qu\'{\i}mica, Universidade de São Paulo,  Av. Prof. Lineu Prestes 748, 05508-900, Butantã, São Paulo, SP, Brazil \\
$^{3}$Observat\'orio Nacional, Rua Gen. José Cristino, 77, 20921-400 - Vasco da Gama, Rio de Janeiro - RJ, Brazil\\
$^{4}$Observat\'orio do Valongo, Universidade Federal do Rio de Janeiro - UFRJ, Ladeira Pedro Ant\^onio 43, 20080-090, Rio de Janeiro, RJ, Brazil\\
$^{5}$Petrobras, Av. Henrique Valadares, 28, 20231-030, Rio de Janeiro, RJ, Brazil
}

\date{Accepted 2024 April 24. Received 2024 April 24; in original form 2023 September 30}

\pubyear{2024}

\begin{document} 

\label{firstpage}
\pagerange{\pageref{firstpage}--\pageref{lastpage}}
\maketitle

\begin{abstract}
Small rocky planets, as well as larger planets that suffered extensive volatile loss, tend to be drier and have thinner atmospheres as compared to Earth. Such planets probably outnumber worlds better endowed with volatiles, being the most common habitable planets. For the subgroup of fast rotators following eccentric orbits, atmospheres suffer radiative forcing and their heat capacity provides a method for gauging atmospheric thickness and surface conditions. We further explore the model presented in a previous paper and apply it to real and hypothetical exoplanets in the habitable zone of various classes of stars, simulating atmospheric and orbital characteristics. For planetary eccentricities {\it e} $\sim$0.3, the forcing-induced hypothetical temperature variation would reach $\sim 80$ K for airless planets and $\sim 10$ K for planets with substantial atmospheres. For Kepler-186 f and Kepler-442 b, assuming {\it e} $\sim$0.1, temperature variations can reach $\sim 24$ K. We also consider habitable exomoons in circular orbits around gas giants within the habitable zone, which suffer radiative forcing due to their epicyclic motion. We study several combinations of parameters for the characterization of planets (mass, eccentricity and semi-major axis) and exomoons (mass, orbital radius, albedo and atmospheric characteristics) for different stellar types. For {\it e} $\sim$0.3, exomoon temperature varies up to $\sim 90$ K, while for $\sim$0.6 variations can reach $\sim 200$ K. Such exomoons may plausibly retain their volatiles by continued volcanic activity fueled by tidal dissipation. Although currently undetectable, such effects might be within reach of future Extremely Large Telescope-class telescopes and space missions with mid-infrared and coronagraphic capabilities.
\end{abstract}



\section{Introduction} \label{sec:introduction}

Since the discovery of the first exoplanets in the 90s \citep{wolszczan1992planetary,mayor1995jupiter}, more than 5600 other planets outside the Solar System were confirmed through different observation techniques (\url{http://exoplanet.eu/}), mainly due to the Kepler space missions \citep{2010Borucki}. Many of them are part of multiple systems, and current and future missions such as the Transiting Exoplanet Survey Satellite (TESS), \citep{huang2018tess, luque2019planetary, gilbert2020first, kostov2020toi, rizzuto2020tess, canas2020warm, vanderburg2020giant, newton2021tess} and the Planetary Transits and Oscillations of stars (PLATO) \citep{2016AN....337..961R} are expected to discover many more. 

Many such planets lie within the habitable zone (hereafter HZ) of long lived stars \citep{2015ApJ...800...99T, 2017Natur.544..333D, 2020MNRAS.493..536D}. Systems such as TRAPPIST-1, with seven Earth-sized planets orbiting an M dwarf star \citep{gillon2017seven}, three of which believed to be inside the HZ \citep{2020ApJ...888..122H}, naturally spark our curiosity on the possibility of such systems harboring life. The existence of life, based on our experience on Earth, depends on the existence of water in the liquid state. So far, based on the forms of life we know of, it seems that a rocky exoplanet with stable suitable temperatures and an atmosphere, compatible with liquid water at the surface, would be the most interesting candidate to look for biosignatures.

The stability of brightness temperature of the portion of a rocky planet's atmosphere whose main heat exchange mechanism is radiative, depends greatly on the stability of the flux of radiation received by its host star. This flux can vary significantly due mainly to two effects: variability of the star's emission and the planet's orbital parameters, mainly its eccentricity $e$. In this paper, we will be concerned with the second case, and will not consider any possible variation on the star's radiative emission. Indeed, as can be readily checked on public databases such as the NASA Exoplanet Archive, the mean eccentricity of exoplanets is 0.16, from 2231 confirmed exoplanets with measured values of $e$ in April 16$^{th}$, 2024, while the mean eccentricity of the planets in the Solar System is around 0.047. Therefore, the planets of the Solar System seem to have particularly low values of $e$, while exoplanets present, on average, higher values. Thus, rocky planets with large orbital eccentricities should be fairly common.

As shown in \citet{pinotti2013most}, the presence of an atmosphere contributes to buffer brightness temperature variations due to eccentric orbits, in planets which rotate fast enough so that temperature variations between night and day time can be disregarded. The fraction of the atmosphere which responds radiatively will have a thermal capacitance, and its buffering property will eventually be sensed at the surface level, whose temperature takes into account the greenhouse effect. The brightness temperature will coincide with surface temperature only for the case of no atmosphere. This attenuation and its associated phase lag in relation to the stellar flux profile could, in principle, be detectable through infrared photometric time series, allowing an estimate of atmospheric thickness, as well as, possibly, other atmospheric properties such as gross composition. Therefore, the modeling of such temperature variations can be effective in characterizing the surface conditions of rocky exoplanets when future instrumentation enables the measurement of their mid-infrared emission, through coronagraphic or interferometric techniques, for example Extremely Large Telescope (ELT) / Mid-infrared
ELT Imager and Spectrograph (METIS) \citep{2018SPIE10702E..1UB}. 

The aim of this paper is two-fold: a wider exploration of the model presented in \citet{pinotti2013most} and its application to various sets of real and hypothetical systems. We study the stability of the brightness temperature on different rocky exoplanets, exploring a wide parameter space in both stellar and planetary properties, and discussing on how a periodic temperature profile measurement could be used to indicate the presence of an atmosphere, and to infer some of its main characteristics as well as help establish if a rocky planet is able to maintain liquid water at its surface.

We also study brightness temperature variations for hypothetical rocky exomoons orbiting Jupiter sized exoplanets inside what would be the HZ for a rocky planet. Planetary migration is a common phenomenon, deduced  from their frequently very small semi-major axis \citep{refId0,2019ApJ...884..178W,2009ApJ...691.1764M,Ida_2008}. Volatile-rich moons are expected to lose most if not all of their volatile content as they migrate to the HZ, since their low mass is not capable of holding on to substantial atmospheres.  Another contributing factor is the extended pre-main sequence and early main-sequence phase of higher luminosity that affects low mass stars before they settle onto the main sequence. This extended phase would keep exoplanets and their moons within conditions of rapid water loss for many times $\sim$10$^{\rm 8}$ years and may be relevant for stars as massive as K5 \citep{Ramirez_2014}. For exoplanets about the size of the Earth, complete water loss ensues (unless the planet is endowed with a very large initial water inventory) and it is to be expected that exomoons of giant exoplanets, close to or within the HZ, will suffer a similar fate \citep{2015NatGe...8..177T,2015AsBio..15..119L}. However, for exomoons tidal dissipation as they pursue their fast orbits, maybe tied to relationships with other moons as well, might keep the moon's internal engine going on long and strongly enough to enable protracted volcanism, atmosphere replenishment and crustal rejuvenation.
It is thus to be expected that massive, rocky or partly rocky exomoons might retain a measurable atmosphere, even after migrating inward from their original place of formation at or beyond the snowline. The model of \citet{pinotti2013most} has been upgraded in order to take into account the convolution of the orbit of an exoplanet and its exomoon, and it proposes that the determination of their brightness temperatures could be instrumental in characterizing the exomoons' surface conditions and habitability for a wide range of realistic conditions. Although not yet detected, exomoons are expected to be abundant and are already a popular debate subject among the recent literature \citep{2020A&A...638A..43R, 2021PASP..133i4401D, 2021MNRAS.502.5292Z, 2021MNRAS.501.2378F}, including the recent suggestion of an ``exomoon'' orbiting a nearby brown dwarf \citep{Faherty_2024}. Large rocky exomoons with an atmosphere could be as likely to harbor life as a similar sized exoplanet \citep{Heller_2014,Guimar_es_2018}. In fact, for the case of compact planetary orbits, where tidal locking between planet and star is thought to be a possible cause of significant temperature variations on the planet's surface \citep{Hammond_2021,Eager_Nash_2020}, should not be a problem for the stabilization of the climate on an exomoon, provided that the orbital periods remain not too far from a few terrestrial days, so they can be considered fast rotators.

The paper is divided as follows. In Sec. \ref{sec:methods}, we present the modeling of radiative interaction between a star and the upper atmosphere of a planet in an eccentric orbit and extend it to the case of exomoons. In Sec. \ref{sec:earth}, we present results of the model applied to hypothetical planets using Earth as a base case, in order to demonstrate the net effect on the planet's temperature profile of the different parameters in our model. In Sec.\ref{sec:exoplanets}, we present the results of the model applied to two rocky exoplanets, inside the HZ of their stars and in Sec. \ref{sec:exomoons} we present the results for putative exomoons. In Sec. \ref{sec:tidaldissipation} we discuss estimations of tidal dissipation for large hypothetical exomoons as compared to Solar System cases. We draw our conclusions in Sec. \ref{sec:conclusion}.

\section{Temperature variation model} \label{sec:methods}

Throughout this paper we will use the term brightness temperature with the meaning of ``effective radiating temperature'', since we consider the portion of the atmosphere under investigation as a blackbody, which is a reasonable approximation. 

The  following  mathematical  development  is  based  on  \citet{pinotti2013most},  where  its  details can be appreciated. It considers  the  atmosphere of a planet or a moon as divided between an  outer  part,  whose  main  heat  exchange  mechanism  is  radiative,  and  an  inner  part, dominated  by  convection. The thermal  radiation  of  the  outer  part  is detectable  through  its  brightness  temperature,  $T$. For  a  planet  or  a  moon  subjected  to temporal variation of the stellar radiation flux, $F_{\star}$, the brightness temperature responds according to
\begin{equation} \label{eq:model}
    \chi \,c_p\,\frac{dT}{dt} = \frac{F_{\star}(t)\,(1 - A)}{4} - \varepsilon\,\sigma \hspace{1pt}T^4,
\end{equation}
where $A$ is the Bond albedo, $\sigma$ is the Stefan-Boltzmann constant, $\varepsilon$ is the planet's or moon's emissivity, which is close to 0.9 in the IR \citep{DePaterImke2001Ps}, $c_p$ is the  atmospheric mole heat capacity, and $\chi$ is the mole column density of the atmospheric section considered, which in turn is a fraction of the total atmospheric thickness (only the outer part with brightness temperature $T$). 
The column density is related to the scale height $h$ by the relation ($\chi=h\,\rho$), where $\rho$ is the average mole density. 

Notice that, for the case of no atmosphere ($\chi =0$), we recover from Eq. (\ref{eq:model}) the well-known relation between flux and temperature:
\begin{equation} \label{eq:model_noatm}
    F_{\star}(t)\,(1 - A) = 4\,\varepsilon\,\sigma \hspace{1pt}T^4.
\end{equation}

For a circular orbit, $dT/dt = 0$, $F_{\star}$ is constant and Eq. \ref{eq:model_noatm} is valid even for a planet with atmosphere. In this case, however, no information about $\chi$ can be retrieved by measuring $T$.

For an elliptical orbit, $F_{\star}(t)$ will be a periodic function, and $T(t)$ will respond periodically, showing attenuation (reduction of amplitude) and phase lag, due to the atmospheric heat capacitance, relative to the case of no atmosphere ($\chi=0$). For a sufficient number of observations of $T(t)$, and using Eq. \ref{eq:model}, estimates of $A$ and $\chi$ could be made.

Several limitations of the model must be kept in mind. Firstly, Eq. (\ref{eq:model}) assumes that the planet or moon is a fast rotator, that is, the radiative timescale of the atmosphere is larger than the rotation period. Secondly, the model treats radiation bolometrically, in a simplified way, with no regard to the details of the radiative transfer and its spectral distribution. Thirdly, it further assumes that the planet or moon is volatile deficient so that energy transfer in absorption, evaporation and condensation does not have to be taken into account, and the albedo, which relates to a cloud coverage, which in turn depends on volatile content in the surface of the planet or moon, does not fluctuate considerably along an eccentric orbit. The model also works if a planet rich in volatiles has an average surface temperature high enough so that the volatiles are mostly in the gaseous phase. While the problem of the origin of the Earth’s oceans remains unresolved, with arguments for both wet-endogenous and exogenous sources \citep{van_Dishoeck_2014,2019A&A...622A.208D,2019A&A...624A..28I}, many researchers indicate that prolonged exposure to XUV radiation and particle winds, especially for planets around low-mass stars, may promote substantial volatile loss \citep{2016A&A...596A.111R, 2017ApJ...836L...3A, 10.1093/mnras/stw2578} from rocky planets.

Furthermore, rocky planets inside the HZ of M stars may also be intrinsically water deficient from the point of view of wet-endogenous source (\cite{Lissauer_2007}, but see, for example, \cite{2017A&A...598L...5A} for alternative scenarios). More severe volatile loss is expected for low mass planets, due to their lower gravitational fields, and also due to their weaker magnetosphere \citep{2013ApJ...770...23Z}. Therefore, Eq. \ref{eq:model} applies best to planets or moons smaller than the Earth, for which volatile scarcity and thin atmospheres should be a common condition. Even planets as massive as the Earth or even larger may frequently incur in severe atmospheric loss and thus be describable by our model. A more recent discussion on volatile loss and water origin in rocky planets can be found in \citet{10.1093/mnras/staa3260}.

Studies  of  planetary  atmospheres subjected  to  radiative  forcing  from  their parent  stars,  and  with  sophisticated  tools  like  Global  Circulation  Models  \citep{Palubski_2020,827206c76f7a4e67abd83a9972b430d2} are usually focused on Earth-like planets, rich in volatiles, having, therefore, more complex energy fluxes, including water evaporation and condensation. Besides, the focus of those studies is on habitability conditions only. Our work, on the other hand, is devoted mainly to an observational method for estimating the atmospheric thickness of planets or moons with low volatile content, which may well be the majority in the Universe.

The physical effect described by Eq. \ref{eq:model} is strengthened as attenuation and phase lag grow. Attenuation depends strongly on the frequency of the periodic forcing, in this case, on $F_{\star}(t)$. In order to appreciate this effect, let us first analyze the behavior of $T(t)$ for a planet in an almost circular orbit where $F_{\star}(t)$ can be approximated as:
\begin{equation} \label{eq:planet_forcing}
F_{\star}(t) \simeq F(a) + F(a)\, \left[\frac{1}{(1-e)^2}-1 \right]\,{\rm sin} (\omega\,t),    
\end{equation}
where $a$ is the planet's semi-major axis, $e$ is its eccentricity, and $\omega$ is the orbital frequency, which, according to Kepler's third law, can be written as:
\begin{equation}\label{eq:omega}
\omega = \frac{(M_{\star}\,G)^{0.5}}{a^{1.5}},
\end{equation}
where $G$ is the gravitational constant, and $M_{\star}$ is the mass of the star. 

Now, for a sinusoidal forcing of the form of Eq. \ref{eq:planet_forcing}, the response of $T$ in a linearized form of Eq. \ref{eq:model}, using Taylor series around a point at $T(a)$ with stable near circular orbit, is:
\begin{equation}
\bar{T} = K\,H\, \left(\omega^2\,\tau^2 + 1 \right)^{-0.5}\,{\rm sin}\,(\omega\,t + \phi),
\end{equation}
where $\bar{T}=T-T(a)$, $K$ is the system gain ($K=[4\,\sigma\, \varepsilon(T(a)^3)]^{-1})$, $H$ is the amplitude of the forcing, $\tau$ is the time constant of the linearized form of Eq. \ref{eq:model} (see \cite{pinotti2013most}), and $\phi$ is the phase lag ($\phi=-{\rm tan}^{-1}(\omega\,\tau$)). We see then that the amplitude of $\bar{T}$ is affected by the term $(\omega^2\,\tau^2 + 1)^{-0.5}$, that is, the attenuation or damping of the amplitude of $\bar{T}$ is related to $\omega$, which in turn depends strongly on the value of $a$ (see Eq. \ref{eq:omega}). That is why the damping will be more evident for planets in the HZ of later types stars. The attenuation is also important in terms of habitability, and this relation is discussed in \citet{pinotti2013most}.

Now let's consider the case of a moon (with a substantial atmosphere) and its parent planet, where the orbit of the moon around the planet is circular, and the orbit of the planet around its star is elliptic. In this case, the form of $F_{\star}(t)$ will be the result of two superimposed movements. Considering that the radius of the moon's orbit around the planet is $r << a$ and the above considerations of simplification (linearization) of the main equation in order to allow a clearer view of the entire system, we can approximate $F_{\star}(t)$ as:
\begin{equation} \label{eq:moon_forcing}
F_{\star}(t) \simeq F(a) + F(a)\, \left[\frac{1}{(1-e)^2}-1 \right]\,{\rm sin} (\omega\,t) + \left[F(a) - F(a-r) \right]\, {\rm sin}\,(\omega_2\,t),
\end{equation}
where $\omega_2$ is given by:
\begin{equation}\label{eq:omega2}
\omega_2 = \frac{(M_{p}\,G)^{0.5}}{r^{1.5}},
\end{equation}
and $M_p$ is the mass of the planet. The response of the moon's brightness temperature to the forcing is:
\begin{equation}
\bar{T} = K\,H\, \left(\omega^2\,\tau^2 + 1 \right)^{-0.5}\,{\rm sin}\,(\omega\,t + \phi) + K\,J\,(\omega_2^2 \tau^2 + 1)^{-0.5}\, {\rm sin}\, (\omega_2\,t + \phi_2),
\end{equation}
where $J$ is the amplitude of the second sinusoidal forcing caused by the revolution of the moon around the planet and $\phi_2=-{\rm tan}^{-1}(\omega_2\,\tau$). That is, the response is a convolution of the individual responses to each periodic forcing. The relative strength of each individual response depends on the orbit parameters, but it is expected that $J << H$, since
\begin{equation}
F(a) - F(a-r) << F(a)\, \left[ \frac{1}{(1-e)^2} - 1 \right].
\end{equation}

The above mathematical development allows the study of a wide spectrum of orbital configurations and their corresponding effects on $T$. For example, if the orbit of the planet is circular ($H = 0$),  the temperature response will be  due  to  the moon's orbit around the planet only. Also, if the planet is far enough from the star, $J$ may be too  small  for  any  effect  on  the  brightness  temperature  to  be  detectable. It is  also  expected that $\omega_2 >> \omega_1$, so the general picture of a $T$ versus time plot for a moon should look like a long periodic function, ``dented'' by the influence of the orbit of the moon around the planet. However, for planets around M dwarfs, $\omega_1$ and $\omega_2$ may become similar.  For the particular case of a circular orbit and an airless moon there will be no attenuation on the oscillations, since $\tau = 0$. In this case, the term $F(a - r)$ will be responsible for the amplitude of the temperature variations and higher relative contributions from close stellar orbits will produce larger variations. In the case of moons with atmospheres, this effect will be balanced by the opposite effect of the attenuation while the resulting temperature variation amplitudes will depend on their individual magnitudes. We will explore the interaction between exoplanet's and their exomoon's orbit in Sec. \ref{sec:exomoons}.

\section{A hypothetical Earth: a case study} \label{sec:earth} 

In order to further understand our model's behavior and its dependence on each free parameter, we applied it to hypothetical Earth analogs. For these hypothetical Earths, we solved Eq. (\ref{eq:model}) while varying the value of each parameter individually and then studied their influence on the temperature profile through four indicators: attenuation, phase lag (both already studied in \citet{pinotti2013most}), asymmetry, and narrowing, proposed in the current paper.

As stated in Sec. \ref{sec:methods}, our model disregards intrinsic stellar flux variations and estimates a planet's brightness temperature by using only its orbital eccentricity, albedo, emissivity, atmospheric column density (dominated by radiation heat transfer), and heat capacity. These are the five parameters we will consider here.

\begin{figure*}
\centering
\includegraphics[width=\linewidth]{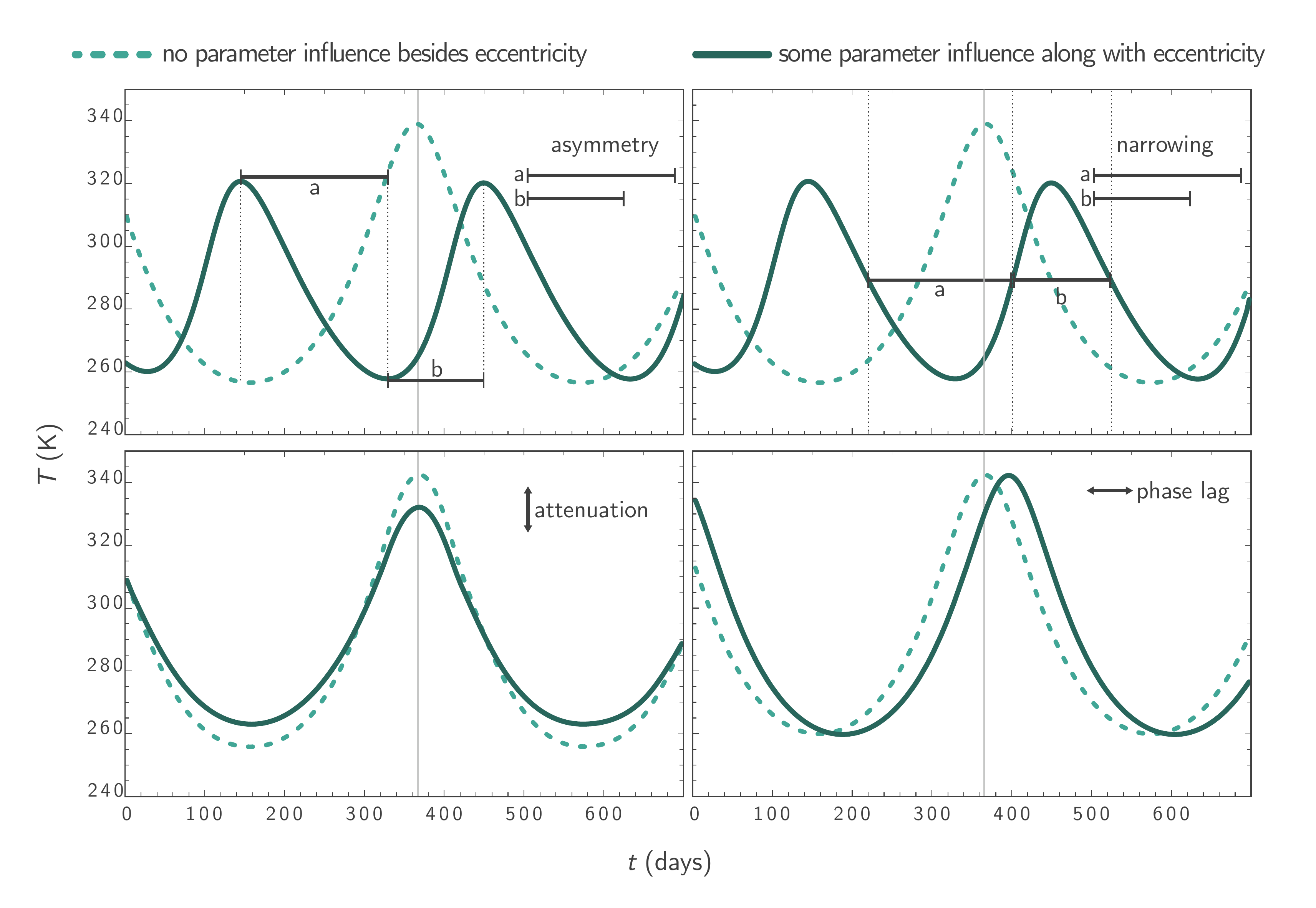}\vspace{.1cm}
\caption{Illustration of the four indicators used in this work on the analysis of the curves, such as attenuation, phase lag, asymmetry, and narrowing. In these illustrations, we used a cosine function as a model curve (dashed lines) due to its symmetry to represent a planet without any indicator effect. The solid lines represent the alteration of the temperature variation profile of a planet caused by the parameters studied.}
\label{fig:medidas}
\end{figure*}

The attenuation indicator marks the difference between the maximum effective temperature of a planet without an atmosphere and the one reached by the same planet if it had an atmosphere. 
The phase lag indicates the difference between the time that the planet reaches maximum brightness temperature in both scenarios (with and without an atmosphere).
The asymmetry indicator proposed in this paper measures the difference between the time taken by the planet to cool when it leaves the periastron towards the apoastron, compared to the time that it takes to heat when it gets closer to its star moving from the apoastron to the periastron.
The indicator that we named ``narrowing'' measures the smoothing of the temperature variation in the apoastron compared to it on the periastron. In other words, it indicates how sudden is the temperature variation amplitude when the planet is closest to the star, compared to when it is furthest the star.  An illustration of the four parameters is exhibited in Fig. \ref{fig:medidas}.

Our hypothetical Earth analogs will be built from a template with the same characteristics of our planet: a star with the same luminosity of the Sun (3.846 $\times$ 10$^{26}$ W), semi-major axis of 1 AU and eccentricity of $e = 0.0167$. We adopted an albedo of $A = 0.3$ and emissivity of $\epsilon= 0.9$. We varied each of the five studied parameters individually from this template. To better distinguish between the effects of each parameter, we started by studying the case of a hypothetical Earth without an atmosphere ($\chi$ = 0).

\begin{figure}
    \centering
    \includegraphics[width=\columnwidth]{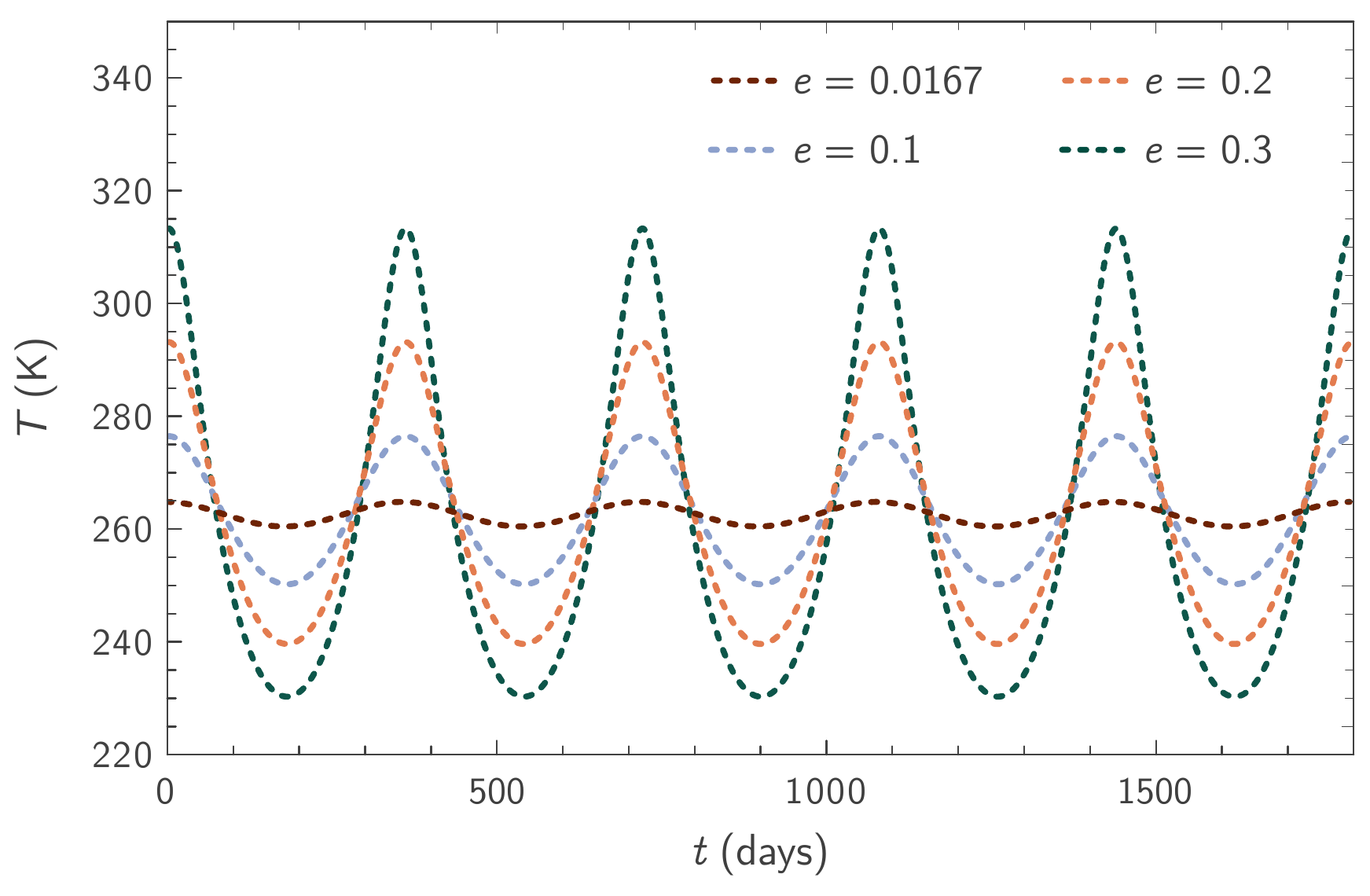}
    \centering
    \caption{Temperature as a function of time for four hypothetical planets without an atmosphere ($\chi = 0$), $A = 0.3$, $\epsilon= 0.9$ and distinct values of eccentricity: 0.0167, 0.1, 0.2 and 0.3.}
\label{fig:earth_ecc}
\end{figure}

Aiming to study how the orbital eccentricity influences the behavior of a planet's brightness temperature, we solved Eq. (\ref{eq:model}) for four distinct eccentricity values: $e=$ 0.0167, 0.1, 0.2, and 0.3. As exhibited in Fig. \ref{fig:earth_ecc}, for highly eccentric orbit, our hypothetical Earth based planet can reach a temperature variation of 80 K throughout the year. Since these planets have no atmosphere, there will be no phase lag or attenuation. One can see that the eccentricity does not cause an asymmetry in the curve, but does cause a narrowing, that increases proportionately: the more elliptical the orbit, the sharper the temperature increase near periastron, a well-known effect that drives Martian global dust storms since southern summer takes place close to maximum proximity to the Sun \citep{SHIRLEY2015128}. The remaining parameters were studied only with the two extreme values of eccentricity (0.0167 and 0.3).

We proceeded with the study of planets without an atmosphere to evaluate how albedo and emissivity impact our results. To evaluate the influence of the albedo for both $e = 0.0167$ and $e=0.3$ values of eccentricity, we solved Eq. (\ref{eq:model}) using two extreme albedo values found in the Solar System, which are 0.1, based on Mercury, and 0.8, based on Venus. As expected, we verified that an increase in the albedo value caused a reduction on the planet's average surface temperature and on the amplitude of its variation throughout the year. In fact, the calculated difference in mean temperature is $\sim$90 K between the two extreme albedos studied, for both eccentricities considered. For the low eccentricity case, there are insignificant (a few K) seasonal temperature variations. For the high eccentricity case, the total temperature amplitude reduces from $\sim$85 K to $\sim$60 K when the albedo changes from 0.1 to 0.8. We studied the effect of changes in planetary emissivity for the same two orbital eccentricities, for the cases of $\varepsilon = 0.7$ and $\varepsilon = 0.9$ and verified that, as expected, the lower the emissivity, the higher the global temperature gets. In fact, the average temperature is reduced by $\sim 20$ K when we change $\varepsilon$ from 0.7 to 0.9 (from $\sim$290 K  to $\sim$270 K for $e=0.3$ and from $\sim$280 K to $\sim$260 K for $e=0.0167$). Temperature variation amplitudes slightly increase for a lower value of $\varepsilon$, but only by a few degrees. Throughout this paper, we will fix $\varepsilon = 0.9$ which, according to these results, is a conservative choice in terms of detectability. Neither albedo nor emissivity introduce an asymmetry in the temperature profiles. However, we verified that both of them introduce a narrowing effect. While the highest albedo value produces a slight narrowing, both emissivity values considered produce a narrowing in the curves which is comparable to the one produced by the eccentricity. The higher the emissivity, the more pronounced this effect becomes.

In order to study the influence of the atmospheric heat capacity and column density, we included an atmosphere to our model, with template values of c$_p$ = 32 J mol$^{-1}$ K$^{-1}$ and $\chi$ = 3 $\times10^{4}$ mol m$^{-2}$. The atmospheric column density value is the same as Earth's, obtained by assuming the radiative part of the atmosphere to start above a pressure value of 0.1 bar \citep{Robinson_2013} and exponential profiles for density and pressure, with scale height $h = 8.5$ km for both (taken from NASA's ``planetary fact sheet" webpage\footnote{\url{https://nssdc.gsfc.nasa.gov/planetary/planetfact.html}}). We decided to use this value of heat capacity based on \citet{pinotti2013most} that considers the sensitivity of heat capacity for relevant gases of rocky planets' atmospheres (CH$_4$, O$_2$, N$_2$, NH$_3$, H$_2$O, CO$_2$ and H$_2$S), which vary between 29.1 J mol$^{-1}$ K$^{-1}$ and 37.3 J mol$^{-1}$ K$^{-1}$ for 300 K. We checked this is valid as well for H$_2$ rich atmospheres, as might be the case in planets with intense volcanic activity \citep{Pierrehumbert_2011,Ramirez_2017}, since H$_2$ has c$_p$ = 28.8 J mol$^{-1}$ K$^{-1}$ at 300 K. The heat capacity and the column density are two important variables that can influence the degree of damping of the upper atmosphere's temperature.

\begin{figure}
    \includegraphics[width=\columnwidth]{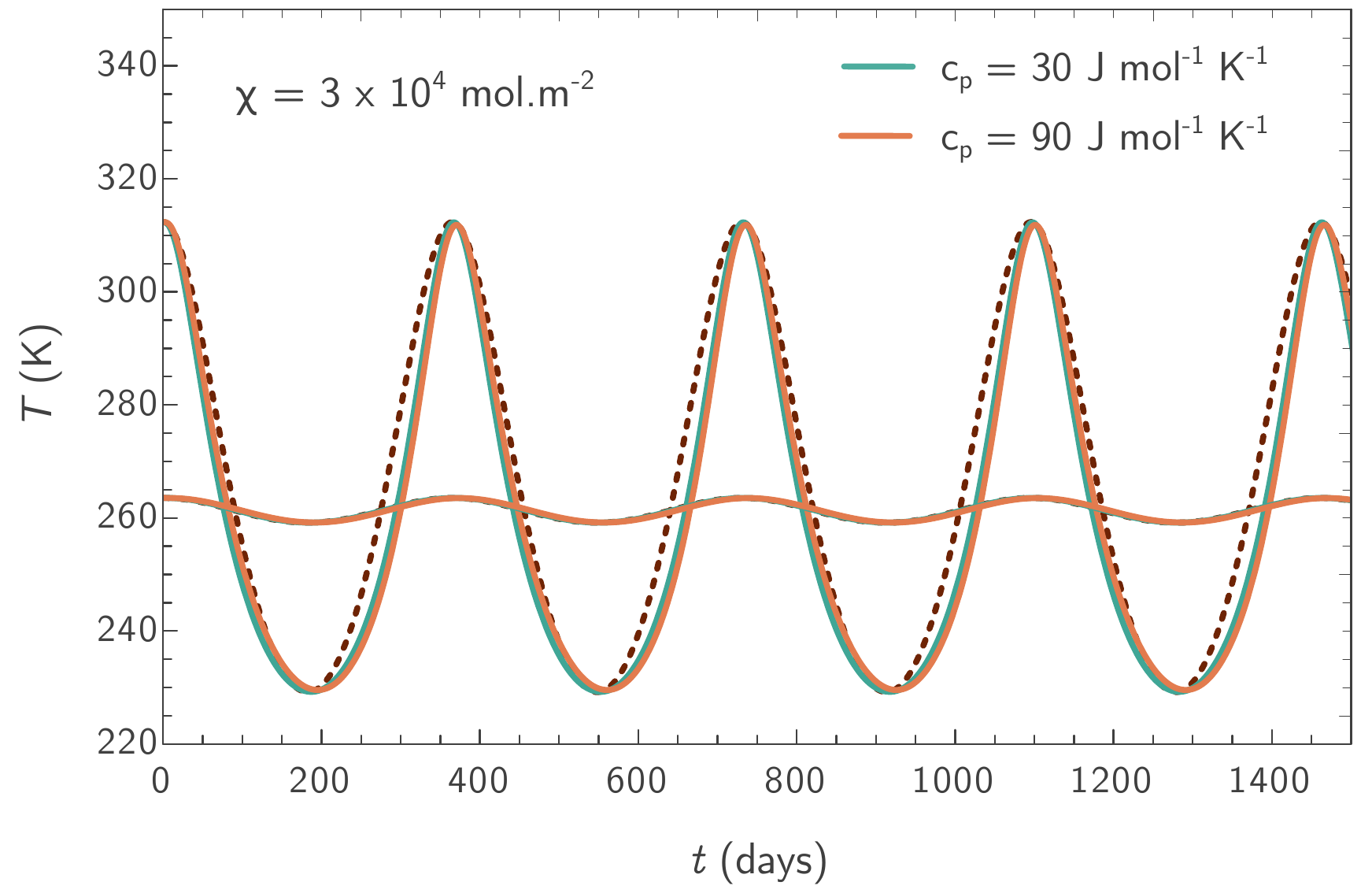}
    \includegraphics[width=\columnwidth]{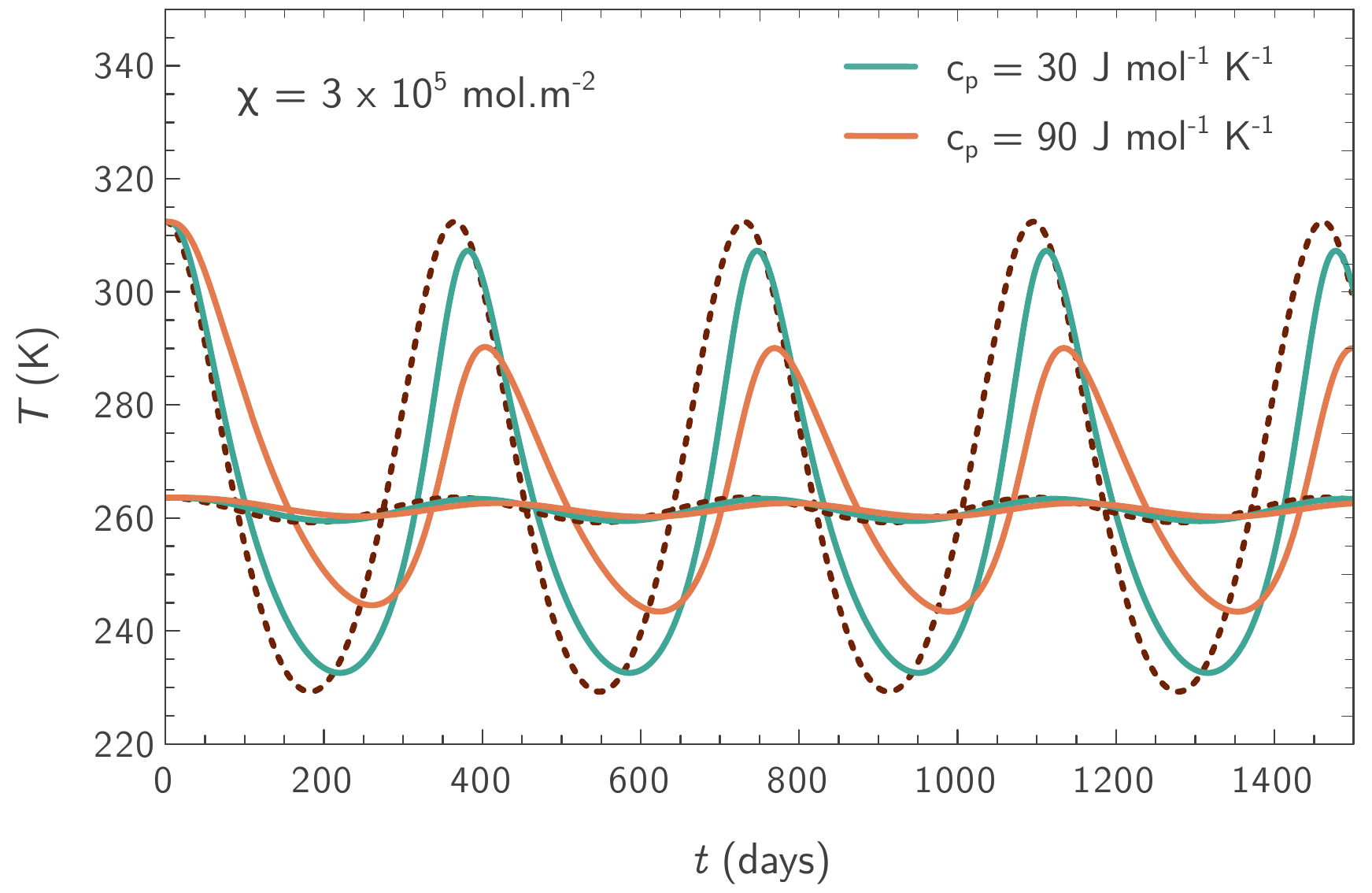}
    \caption{Brightness temperature as a function of time for hypothetical planets with $A = 0.3$ and $\varepsilon=0.9$. The dotted brown curves show the cases with no atmosphere, for comparison, and the solid lines show the cases for an atmosphere with $\chi=3 \times 10^4$ mol m$^{-2}$ for the upper panel and $\chi=3 \times 10^5$ mol m$^{-2}$ for the lower one. The planets have two distinct values of eccentricity: 0.0167 (smallest amplitude curves) and 0.3 (curves with highest amplitude). For the cases with an atmosphere, the heat capacity is either 30 J mol$^{-1}$ K$^{-1}$ or 90 J mol$^{-1}$ K$^{-1}$.}
\label{fig:earth_cp}
\end{figure}

In order to evaluate how our choice of $c_p$ could impact our final results, we considered two very distinct values of heat capacity, 30 J mol$^{-1}$ K$^{-1}$ (which is close to our template choice), as an average of gases that make up the Earth's atmosphere at $\approx$300 K, and 90 J mol$^{-1}$ K$^{-1}$, as an average of the same gases at extremely high temperatures ($\approx$1500 K). We show the results in Fig. \ref{fig:earth_cp}. As can be seen in the upper panel, the two different choices of $c_p$ apparently have no impact for an atmosphere with Earth's $\chi$ value. In order to quantify the impact of $c_p$, we reproduced the same plot in the lower panel, changing only the value of $\chi$ for one 10 times greater. We can see a high heat capacity increases the attenuation, phase lag and asymmetry, which means that the planet suffers a reduction in temperature variation, a slowing not only in the time to reach the maximum temperature but also in its cooling when moving far from its star. This relates to the narrowing effect because the higher heat capacity acted decreasing this indicator, causing the planet to spend a more balanced time with cold temperatures near the apoastron and hot temperatures close to the periastron compared to the lower heat capacity. Since our aim is to study objects inside or near the HZ, a choice of $c_p$ corresponding to a temperature of $\approx$300 K seems more adequate. Also, we can see that even for the very distinct values of $c_p$ studied the effect on the temperature profiles was not as radical as could have been expected, meaning that fluctuations around our template value (c$_p$ = 32 J mol$^{-1}$ K$^{-1}$) would not cause major changes in our results. It should also be noted that the main gases comprising Earth's atmosphere, as well as CO$_2$, have very similar values of $c_p$ and, therefore, $c_p$ is robust against composition variation.

\begin{figure}
    \centering
    \includegraphics[width=\columnwidth]{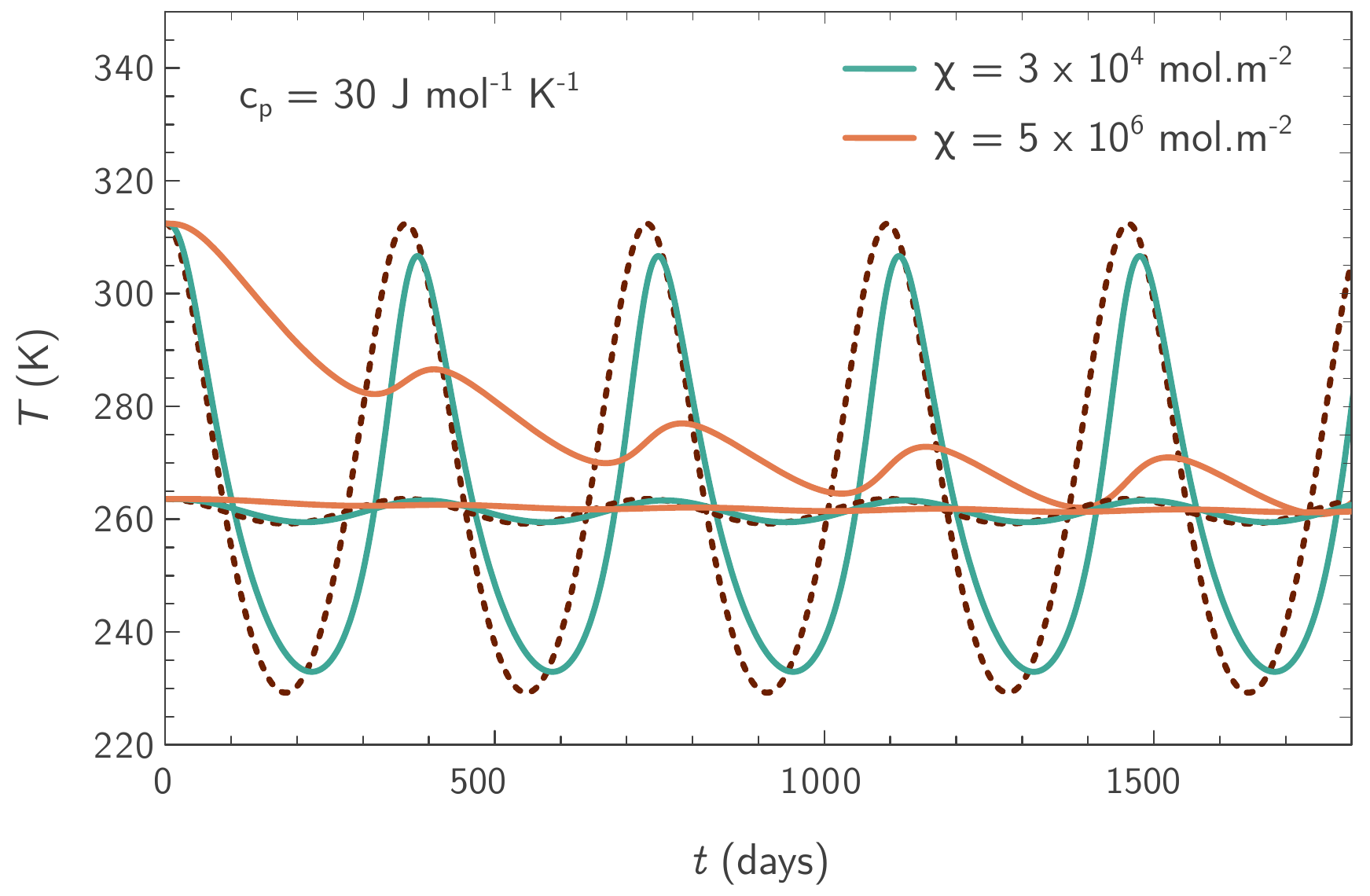}\vspace{.3cm}
    \centering
    \caption{Brightness temperature as a function of time for two hypothetical planets with an atmosphere (solid curves) compared to the corresponding cases without an atmosphere (dotted, brown curves). The planets have two distinct values of eccentricity: 0.0167 (smallest amplitude curves) and 0.3 (curves with highest amplitude) and fixed $\varepsilon = 0.9$, $A=0.3$ and c$_p$ = 32 J mol$^{-1}$ K$^{-1}$. For the cases with an atmosphere, the column density is either $\chi= 3 \times 10^4$ mol m$^{-2}$ or $\chi=5 \times 10^6$ mol m$^{-2}$.}
\label{fig:earth_chi}
\end{figure}

As already verified, the value of atmospheric column density $\chi$ can have a big impact on the temperature profile. In order to further study the influence of $\chi$, we solved Eq. (\ref{eq:model}) using three different values of $\chi$: the first based on Mars' atmosphere ($\chi = 5 \times 10^3$ mol m$^{-2}$), the second based on Earth's ($\chi$ = $3 \times 10^4$ mol m$^{-2}$), and the third one based on Titan's ($\chi = 5 \times 10^6$ mol m$^{-2}$). The values of $\chi$ for Mars and Titan were obtained assuming exponential profiles for density and pressure, with the same value of $h$ for both, as was done for Earth. We considered the thin Martian atmosphere to be fully radiative, with $h = 11.1$ km (taken from  NASA's ``planetary fact sheet" webpage). For Titan, we considered the transition between a convective and radiative atmosphere to occur when pressure reaches 1.3 bar. This value is much higher than the 0.1 bar adopted for Earth (which can also be assumed for other planets) because shortwave absorption in Titan's upper hazy troposphere causes stability against convection \citep{Robinson_2013}. Using values of $h$ between 15 km and 50 km \citep{Horst_2017}, we obtained the average value of $\chi = 5 \times 10^6$ mol m$^{-2}$ for Titan. In Fig. \ref{fig:earth_chi} we can see the temperature profiles for planets with different values of atmospheric $\chi$. We chose to display only the cases for the atmospheres of Earth and Titan, since the results for Mars' atmosphere do not produce any significant changes from the Earth's. Actually, we have verified that, keeping all the other parameters unchanged, significant changes from the temperature profile obtained for $\chi = 3 \times 10^4$ mol m$^{-2}$ (the value for Earth's atmosphere) appear only for atmospheres with $\chi \gtrsim 3 \times 10^5$ mol m$^{-2}$. Fig. \ref{fig:earth_chi} shows the temperature profiles for the two eccentricity cases we have selected compared to the same planet without an atmosphere. All the curves were analyzed with the four indicators. The data shows that the presence of an atmosphere not only strongly attenuates the temperature variation but also causes a phase lag. A thicker atmosphere also increases the asymmetry, stressing the elliptical nature of the orbit, and decreases the narrowing, which causes the curve to be smoother.

\section{Temperature variation of real exoplanets} \label{sec:exoplanets} 

In this section we present the results of the adopted model for real exoplanets which are considered to be inside the HZ of their host stars. We explore the parameter space of albedo, eccentricity and atmospheric thickness. 

\subsection{Selection criteria} \label{subsec:selection}

We have selected two exoplanets from all the confirmed exoplanets available at the NASA Exoplanet Archive Database\footnote{https://exoplanetarchive.ipac.caltech.edu/} that we classified as potentially rocky and are inside the classical HZ of their stars, with the assumptions used in \cite{1993Icar..101..108K} and \cite{kopparapu2013habitable}. We calculated the planetary density for the subgroup of exoplanets that present measured mass values using the results from the interior planetary models \citep{zeng2016mass,zeng2019growth}. We considered as potentially rocky planets those that meet both of the following conditions: planetary density between 4 g/cm$^{3}$ and 10 g/cm$^{3}$ and planetary radius less than $2\,R_{\oplus}$. 

The HZ around the stars were estimated according to the formalism presented in \citet{kopparapu2013habitable}, although other definitions exist in the literature as, for example, in \citet{Ramirez_2017} and \citet{Pierrehumbert_2011}. The effective flux received by the planet (S$_{eff}$) was calculated considering the effective temperature of the star (T$_{eff}$) and the coefficients $a$, $b$, $c$ and $d$, taken from \citet{kopparapu2013erratum} when considering an exoplanet with $1\,M_{\oplus}$. The limits for the inner and outer borders of HZ calculation were chosen as the runaway greenhouse and the maximum greenhouse effect, respectively. 

From this selection, we chose exoplanets Kepler-186 f and Kepler-442 b as targets for our model. Table \ref{tab:exoplanetsparam} shows information for these two exoplanets and their corresponding host stars. These planets are particularly promising cases for our model since they lie much closer to their host stars than the Earth while still within the HZ. Thus, despite being larger than the Earth, they both have been exposed to a far greater XUV fluence than the Early Earth, and may have suffered significant atmospheric loss. Under the assumption that they formed with early atmospheres as massive as the Earth's, their losses may well place them within the reach of our calculations, focused on planets with thin atmospheres.  

We have used a fixed value of eccentricity of $e = 0.1$ in our calculations, which is within the uncertainties for the measured minimum eccentricities ($0.04^{+0.07}_{-0.04}$ and $0.04^{+0.08}_{-0.04}$ for Kepler-186 f and Kepler-442 b, respectively \citep{2015ApJ...800...99T}) and a sizable enough value to produce strong results from our model. We have also simulated different atmospheric thicknesses and albedos, as shown in Table \ref{tab:exoplanetsatm}. Albedo values were chosen to be in rough agreement with the types of atmosphere under consideration and, therefore, we used as reference Mars' and Earth's mean albedos, which are 0.25 and 0.30, respectively \citep{perryman_2018}. We also included the case of albedo $A = 0.40$ to take into consideration possible more icy or cloudy exoplanets. We considered an albedo of $0.27$ to model Titan's atmosphere \citep{NEFF1985425}.

The resulting brightness temperature profiles estimated by our model are mainly affected by atmospheric specifics, but also by eventual stellar flux variations and the existence of other sources of heat, such as tidal heat. Inaccuracies inherent to the model as well as the assumptions made are certainly the chief sources of uncertainty in our results. However, the uncertainties associated with the values of the measured parameters are also sources of error. The values of the physical parameters of the host stars and exoplanets we considered are, in general, determined with relative uncertainties of less than $\sim 10\%$. Measurements of orbital eccentricities are less precise. For the sake of clarity, we will not present uncertainties in our plots, but in the course of the text we will comment on the expected variations in our results due to measurement precision. 

\begin{table*}
	\centering
	\caption{Radius ($R_P$), semi-major axis ($a$), minimum orbital eccentricity ($e_{min}$) and orbital period ($P$) for exoplanets Kepler-186 f and Kepler-442 b and host stars' mass ($M_{\star}$), radius ($R_{\star}$), effective temperature ($T_{eff}$), luminosity ($L_{\star}$) and metallicity [Fe/H]. Data were taken from \citet{2015ApJ...800...99T}.}
	\label{tab:exoplanetsparam}
	\begin{tabular}{cccccccccc} 
		\hline
		Exoplanet & $R_P$ ($R_\oplus$) & $a$ (AU) & $e_{min}$ & $P$ (days) & $M_{\star}$ ($M_{\odot}$) & $R_{\star}$ ($R_{\odot}$) & $T_{eff}$ (K) & $L_{\star}$ ($L_{\odot}$) & [Fe/H] (dex) \\ 
		\hline
	    Kepler-186 f & 1.17 & 0.432 & 0.04 & 129.9 & 0.544 & 0.523 & 3755 & 0.055 & -0.26\\[.2cm]
		Kepler-442 b & 1.34 & 0.41 & 0.04 & 112.3 & 0.609 & 0.598 & 4402 & 0.117 & -0.37\\
		\hline
	\end{tabular}
\end{table*}

\begin{table}
	\centering
	\caption{Atmospheric models considered and corresponding albedo values. Each model is based on the atmospheres of Mars, Earth and Titan.} 
	\label{tab:exoplanetsatm}
	\begin{tabular}{cccc} 
		\hline
		Model & $\chi$ (mol/m$^2$) & Albedo & Type of atmosphere \\ 
		\hline
		1 & $5 \times 10^3$ & 0.25 & Mars  \\
	    2 & $3 \times 10^4$ & 0.3 & Earth  \\
	    3 & $3 \times 10^4$ & 0.4 & Earth with a higher albedo  \\
        4 & $5 \times 10^6$ & 0.27 & Titan  \\
		\hline
	\end{tabular}
\end{table}

\subsection{Results} \label{subsec:results}

\begin{figure}
    \includegraphics[width=\columnwidth]{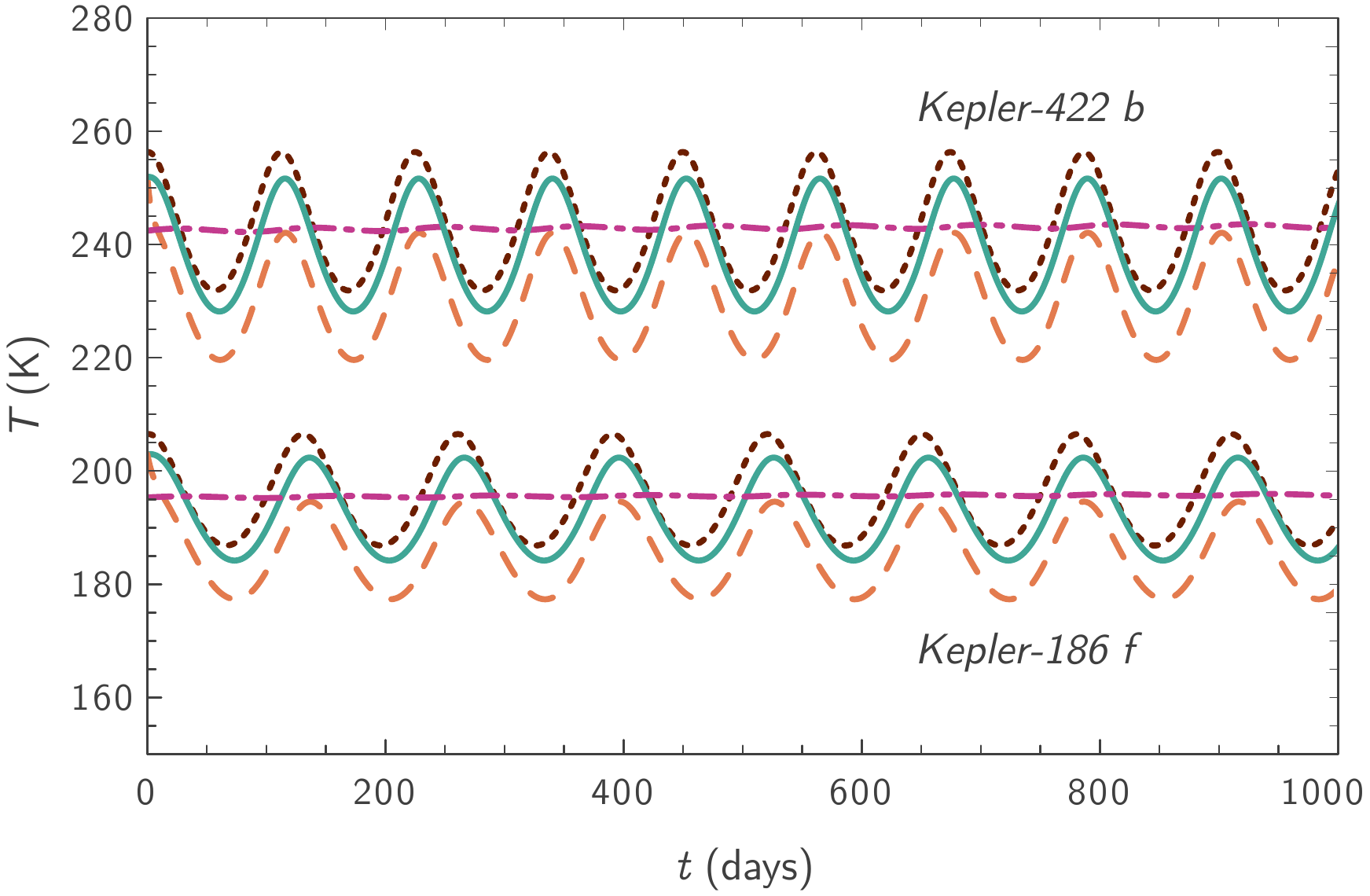}
	\caption{Brightness temperature profiles for Kepler-186 f (lower set of curves) and Kepler-442 b (upper set of curves) for each model in Table \ref{tab:exoplanetsatm}. Dotted brown curves represent model 1, solid curves represent model 2, dashed curves represent model 3 and dot-dashed curves represent model 4.}  \label{fig:RealExoplanets}
\end{figure}

Fig. \ref{fig:RealExoplanets} shows the results of our calculations. The different colors and types of curves represent different sets of parameters for the model. The initial conditions in each case were taken as the temperature at periastron for the exoplanet in the case of no atmosphere (see Eq. (\ref{eq:model_noatm})). 

Models 1 and 2 (see Table \ref{tab:exoplanetsatm}) resulted in similar profiles with offsets given mostly by the different values of $A$. A slight attenuation occurs as the value of $A$ increases, and already mentioned in Section \ref{sec:earth}. 
For the largest value of $\chi$, we clearly see the effect of the presence of an atmosphere on the profiles, which are very attenuated.

For Kepler-186 f, annual temperature variations range from $\sim 0.5$ K, for the thicker atmosphere considered (model 4) to $\sim 20$ K, for the others. Note that, for this planet, stellar flux varies between 300 and 500 W/m$^2$ during one orbit. If we take into account the uncertainties of the measured stellar and planetary parameters (with the exception of eccentricity), these temperature ranges do not vary significantly for model 4, but can vary from $\sim 18$ K to $\sim 24$ K for the other models. For Kepler-442 b, the variations range from $\sim 0.5$ K (from $\sim 0.2$ K to $\sim 1$ K, taking uncertainties into account) to $\sim 24$ K (from $\sim 23$ K to $\sim 33$ K). In this case, the planet goes through a wider range of stellar fluxes (from 800 to 1200 W/m$^2$), since Kepler-442 is brighter and both planets have close values of the orbital parameters. If we account for the fact that the minimum eccentricity can be zero, the orbits would be circular and, of course, temperature variations would be null. 

One can notice that, in all cases, temperatures are below the freezing point of water. This is expected since our model does not take into account the greenhouse effect in the lower atmosphere, whose main heat exchange mechanism is convective. This effect is known to be responsible for a increase of $\sim 35$ K on Earth's temperature \citep{DePaterImke2001Ps}, for example.

\section{A first look at exomoons} \label{sec:exomoons}

In this section, we present the results of our approach when applied to hypothetical rocky exomoons, orbiting giant hypothetical exoplanets belonging to the warm Jupiter class. We also selected some real giant exoplanets and placed hypothetical exomoons orbiting them. Before starting, we performed some tests for Solar System moons, in order to verify the performance of our approach.

In the rest of this section, we model moons in circular orbits around their planets, which is a reasonable approach since they are expected to be in tidal lock. We do not take into account the effect of eclipses, when the planet blocks the stellar flux received by the moon, since we consider that these events will last a negligible amount of time, if compared with the orbital periods considered \citep{2017A&A...601A..91D}. In fact, just like for the Galilean moons of Jupiter, for general orbits aligned at the same plane, eclipses do not affect much the total flux received. We also verified that the center of mass of the planet-moon system lies inside the radius of the planet in all cases, which means that a model of the moons orbiting the center of the planets can be considered as a good approximation. We further neglect the contribution of planetary illumination, since we focus on exomoons within the stellar HZs, for which stellar illumination will strongly dominate over the planetary one, at least for old, evolved systems. Even though this contribution may be relevant in certain cases, it can be reasonably expected not to introduce any serious effect in our results given the scenarios we model \citep[see][]{2017A&A...601A..91D,2017MNRAS.472....8Z}.

\subsection{Tests with Solar System's moons}

In order to show the agreement of our results with actual measurements, we applied the procedure described in Section~\ref{sec:methods} for Titan, Ganymede and Europa. The values used in our model for each moon are shown in Table \ref{tab:solarmoons}. We assumed Ganymede and Europa have no atmosphere and Titan has an atmosphere with $\chi = 5 \times 10^6$ mol/m$^2$ and $c_p = 32$ mol$^{-1}$ K$^{-1}$, as already described in Section \ref{sec:earth}. All moons were assumed to have a circular orbit around the planet and we used for Saturn a mass of $95.16\, M_{\oplus}$, a semi-major axis of $9.583$ AU and an eccentricity of $0.0565$. For Jupiter, we used adopted a mass of $M_J=317.83 \,M_{\oplus}$ a semi-major axis of $5.204$ AU and an eccentricity of $0.0489$. All these values we taken from NASA's ``planetary fact sheet" webpage.\\

\begin{table}
	\centering
	\caption{Mass, semi-major axis, Bond albedo and assumed atmospheric column density for moons in the Solar System. A value of $\chi = 0$ means we assumed the moon has no atmosphere. Values of mass and semi-major axis were taken from the NASA's ``planetary fact sheet" webpage. References for the values of Bond albedo are \citet{NEFF1985425} for Titan, \citet{SQUYRES1982117} for Ganymede and \citet{BURATTI198393} for Europa.}
	\label{tab:solarmoons}
	\begin{tabular}{ccccc} 
		\hline
		Moon & $M$ ($10^{20}$ kg) & $a$ ($10^3$ km) & $A$ & $\chi$ (mol/m$^2$) \\ 
		\hline
		Titan & 1345.5 & 1221.8 & 0.27 & $5 \times 10^6$ \\
	    Ganymede & 1481.9 & 1070.4 & 0.35 & 0  \\
	    Europa & 480.0 & 671.1 & 0.62 & 0  \\
		\hline
	\end{tabular}
\end{table}

\begin{figure}
	\includegraphics[width=\columnwidth]{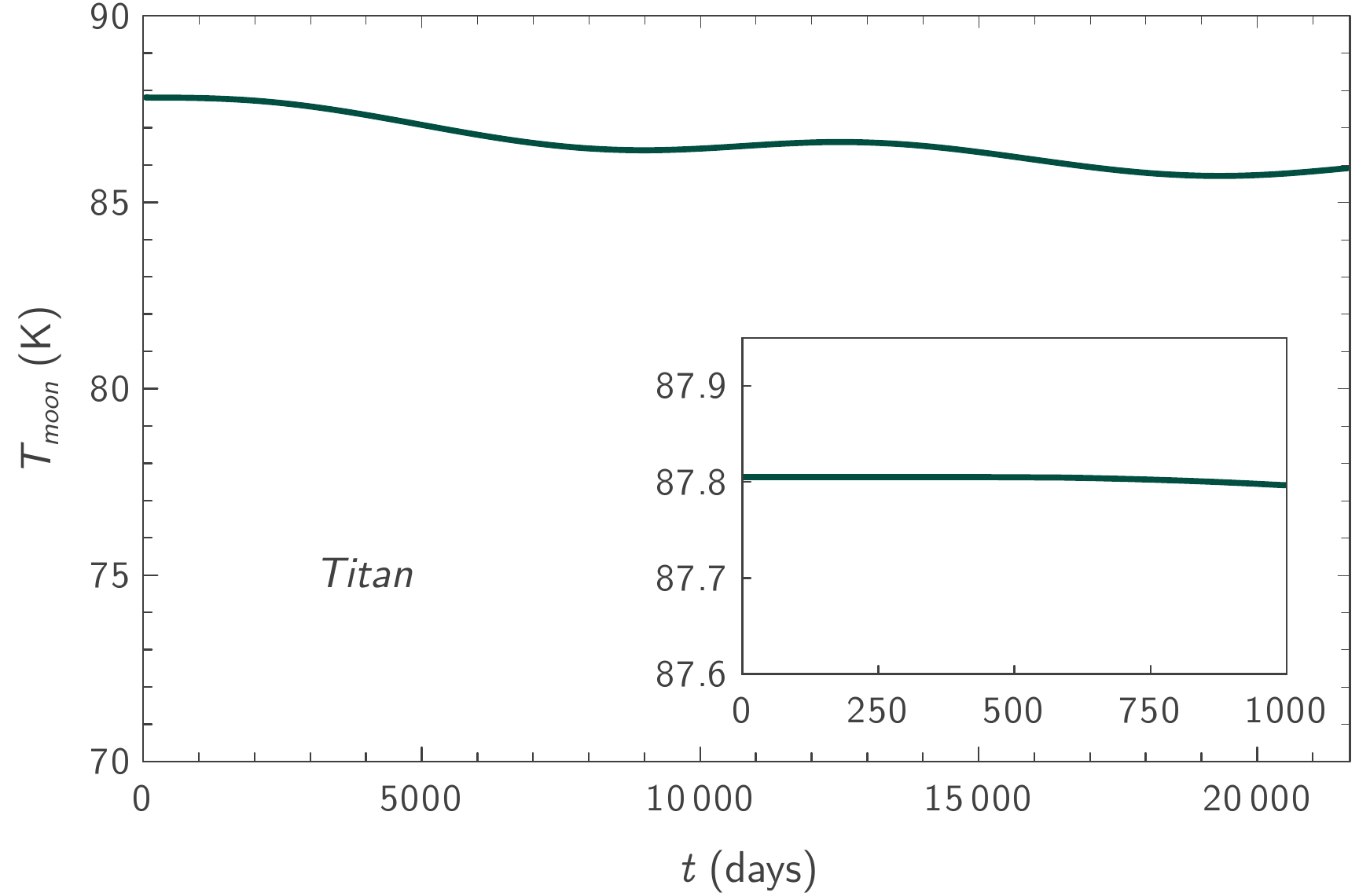}
	\includegraphics[width=\columnwidth]{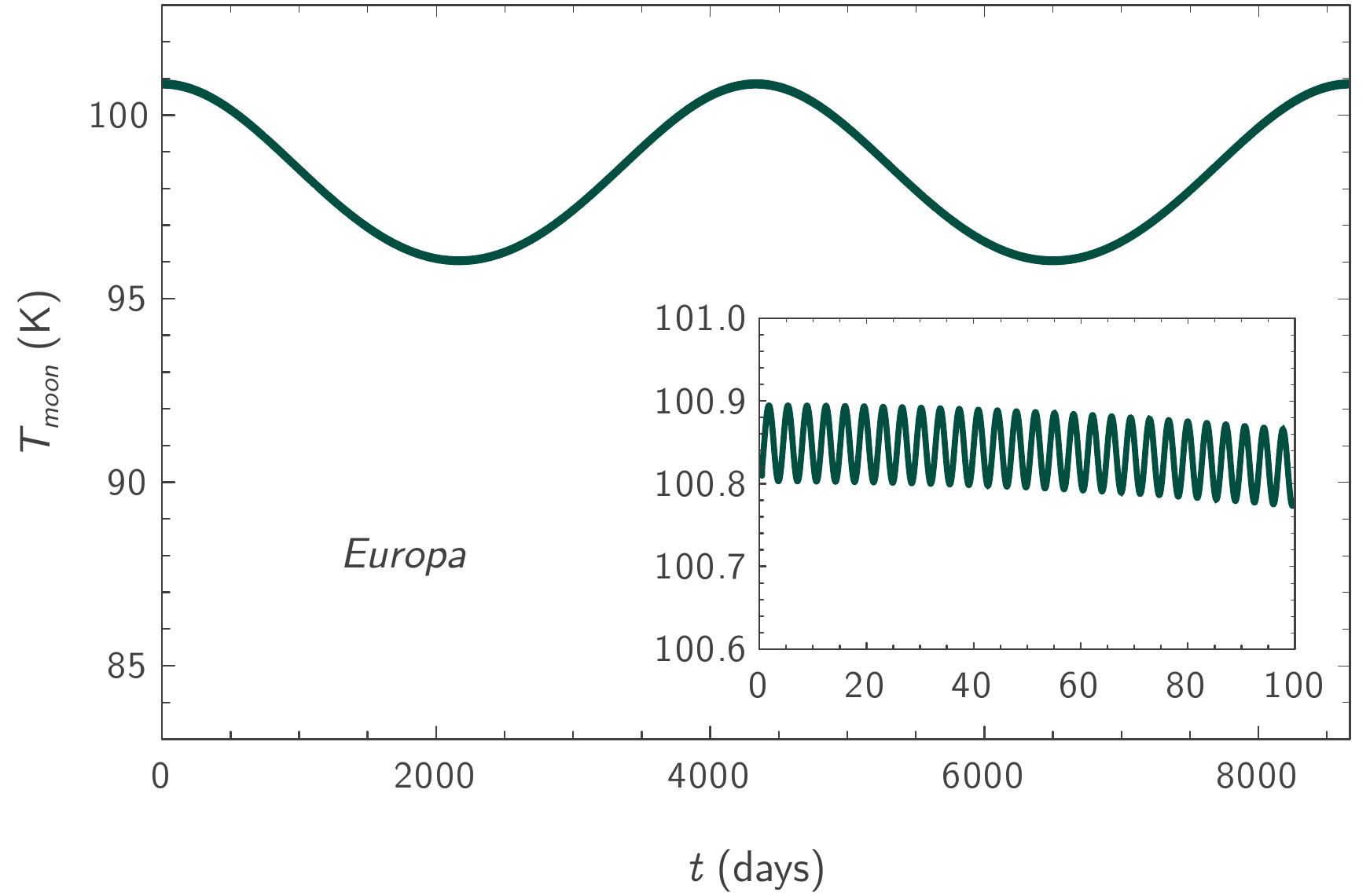}
	\includegraphics[width=\columnwidth]{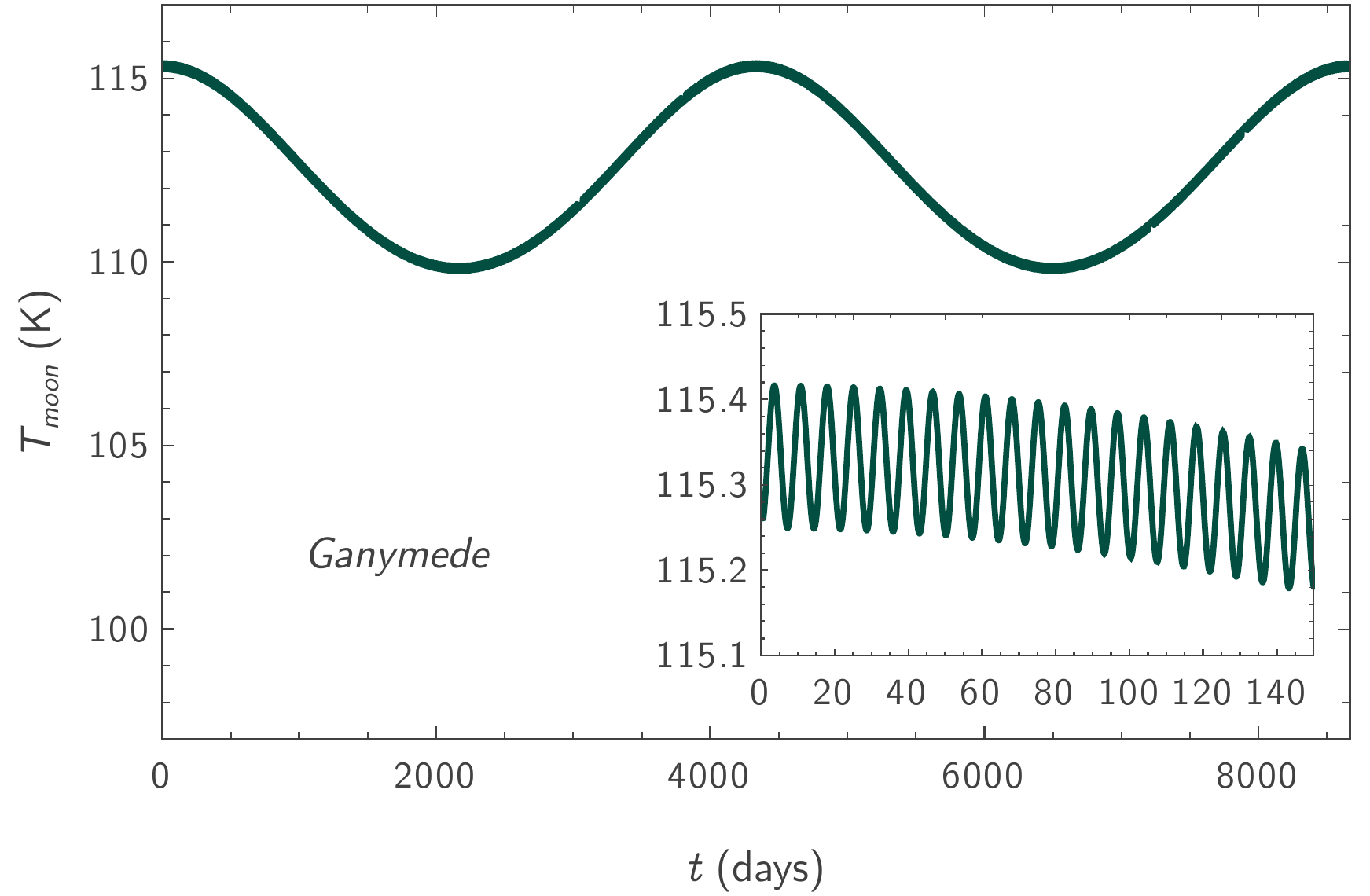}
	\caption{Brightness temperature profile for the three Solar System moons studied. The upper panel shows the result obtained for Titan, the middle one shows the result for Europa and the lower panel for Ganymede. The insets show a zoomed image of the initial part of each curve, so that details of the effect of the moon's orbit  around the planet can be seen.}    \label{fig:SolarSystemMoons}
\end{figure}

The results can be seen in Fig.~\ref{fig:SolarSystemMoons}, in which higher frequency oscillations due to the moons' orbits around the planets are superimposed to lower frequency ones, due to the planets' orbits around the star. The brightness temperature ranges predicted for each moon can be compared with the measured values of the surface temperatures for each one: 89 K $ < T < $ 94 K for Titan \citep{jennings2016surface}, 75 K $<T<$ 152 K for Ganymede \citep{squyres1980surface,orton1996galileo} and 86 K $<T<$ 132 K for Europa \citep{spencer1999temperatures}. Notice that, the brightness and surface temperatures should be the same only for the case of no atmosphere, which may be a good approximation for Europa and Ganymede, but does not apply to Titan. Also, Europa, with a rotation period of $~3.5$ days, is the only moon that can be considered as a fast rotator among the three. Nevertheless, our model produces temperature ranges similar to measurements for these three moons. This lends confidence that our model may realistically represent cases for which conditions better approach our assumptions.

For Ganymede and Europa, we can see the two superimposed oscillations, as expected from our model. In the case of Titan, however, the effects of the its orbit around Saturn does not appear in the temperature profile. This is due to its thick atmosphere, which attenuates temperature variations.

In this case, as it is expected, temperature variations seen in Fig.~\ref{fig:SolarSystemMoons} are essentially due to the slight eccentricity of the planet's orbit, and the second order effects arising from the moons' orbits are negligible, since the distances of the host planets from the Sun are large. In the case of exomoons orbiting giant planets within the HZ, however, we expect the effect of the moons' orbits to be more significant.

\subsection{The case of circular planetary orbits} \label{sec:circular}

Let us now analyze the temperature variations on the surface of a hypothetical exomoon in a circular orbit around a planet which is also in a circular orbit around its host star and within its HZ. In this case, the first order effect due to planetary eccentricity will not be present and temperature variations will be caused by the epicyclic nature of the moons' orbits alone. We will consider two template cases of connected properties between the planet and its moon, and the host star:

\begin{enumerate}
    \item a moon with the mass of Ganymede orbiting a planet with the mass of Jupiter, which orbits an M dwarf star;
    \item a moon with the mass of Mars ($0.107\,M_{\oplus}$) orbiting a planet with three times the mass of Jupiter, which orbits an F dwarf star.
\end{enumerate}

This choice of parameters likely represents extremes, and should cover a wide range of real cases. Jupiter-sized planets are scarce around red dwarfs \citep[see, for example,][]{Kurtovic_2021}, and a Ganymede-sized satellite is probably as large an exomoon as can be expected. In fact, it seems that more massive and larger planets are likelier to form around more massive stars \citep{Lozovsky_2021} and also to host larger and more massive moons \citep{2020MNRAS.492.5089M,
2022MNRAS.513.5290D}. A Mars-sized moon (four times as massive as Ganymede) is a reasonable assumption \citep{Tiane2101155118}, though such an object will not necessarily resemble the composition of Mars, and can be expected to have a more mixed rock/ice makeup.

For each of these two extreme cases, we study three possible scenarios: the scenario where the moon has no atmosphere and an albedo of 0.25; a moon with a Mars like atmosphere with $\chi = 5 \times 10^3$ mol/m$^2$ and albedo 0.25 (corresponding to model 1 in Table \ref{tab:exoplanetsatm}); a moon with an Earth like atmosphere with $\chi = 3 \times 10^4$ mol/m$^2$ and albedo of 0.3 (model 2 in Table \ref{tab:exoplanetsatm}). We chose the stars Gliese 687 and HD 115383 to represent the classes of M dwarfs and F dwarfs, respectively. Table \ref{tab:exomoons} shows the parameters adopted for these stars. 

The radii of the planet's orbits are chosen so that they receive from the star the same flux received by Earth from the Sun. They are 0.14 AU and 1.44 AU, for cases (i) and (ii), respectively. The radii of the moon's orbits are chosen in order that they have a period of 3 days around each planet, so they can be considered as fast rotators. This results in radii of 0.004 AU and 0.006 AU, for cases (i) and (ii) respectively. We have determined the Roche radius for these hypothetical planets and exomoons assuming that the moon with the mass of Ganymede has also Ganymede's density and the moon with the mass of Mars has Mars' density. The results are $0.001$ AU for both cases, so we can be assured that their orbital radius lie beyond the Roche limit. The Hill radii are estimated to be 0.009 AU and 0.01 AU, for cases (i) and (ii), respectively, so the satellites lie within the stability zones of their planets \citep{domingos2006stable}.

\begin{figure}
	\includegraphics[width=\columnwidth]{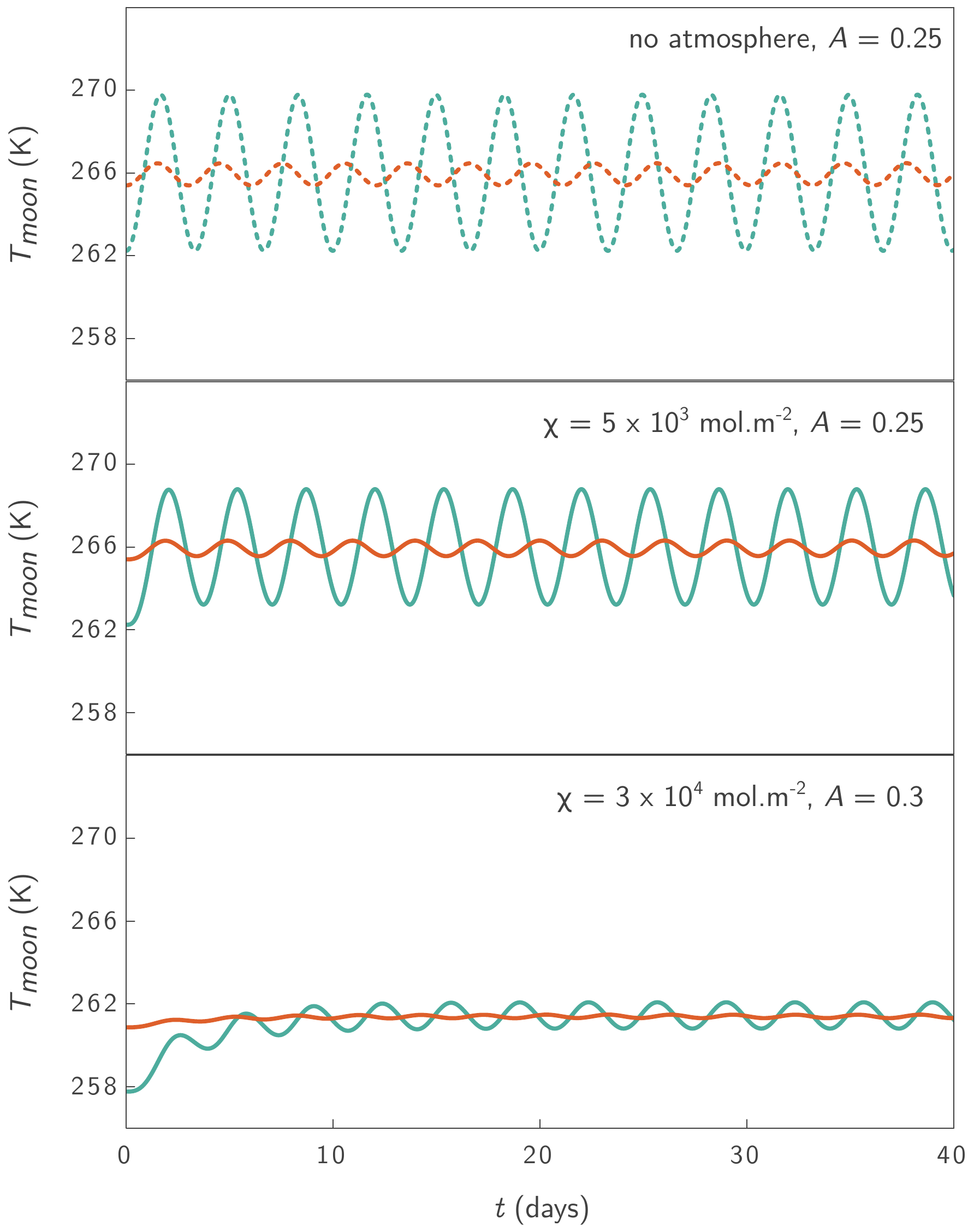}
	\caption{Brightness temperature profile for the hypothetical exomoons studied in Section~\ref{sec:circular}. For each panel, curves with the highest amplitudes are obtained for the lower mass moon (Ganymede’s mass), and curves with
    the lowest ones are obtained for the more massive moon (Mars’ mass). Each panel presents the results for the type of atmosphere and albedo indicated.}   \label{fig:CircularOrbits}
\end{figure}

Fig. \ref{fig:CircularOrbits} shows the resulting temperatures for each case. For the case of a moon with no atmosphere, it is possible to verify that the temperature variations are higher for the system around the M dwarf star, as predicted in Section \ref{sec:methods} for circular orbits. Even when we admit an atmosphere like Earth's, it does not produce a high enough attenuation to compensate for the relative effect of the star's proximity and temperature variations remain higher for the M dwarf case. Temperature variations range from $\sim 0.2$ K for the case of a Mars sized moon with $\chi = 3 \times 10^4$ mol/m$^2$ to $\sim 7-8$ K, for the case of a moon the size of Ganymede with no atmosphere, which is comparable to the results obtained for the Solar System moons.

We have verified that in the case for $\chi = 10^5$ mol/m$^2$ and $A=0.3$ (not shown) temperature variations are heavily damped for both cases of moons and the thermal amplitude is less than $\sim$ 0.5 K. 

\subsection{Eccentric planetary orbits} \label{subsec:eccentricorbits}

Let us now study the case of a hypothetical exomoon in circular orbit around a planet with eccentric orbit around its star. We studied a moon with the mass of Mars ($M_{{\rm exomoon}}=0.107 \, M_{\oplus}$) in a circular orbit around a planet with mass of three times that of Jupiter, which was the case found to yield the most conservative results (smallest temperature variations) in the previous section. The radius of the moon's orbit is determined in order for its period to be of 3 days. This results in an orbital radius of $0.006$ AU, which is $\approx 0.8$ times that of Ganymede's semi-major axis, for example. The Roche limit for this case is 0.001 AU. In this section, we studied the case of four different stellar types, adding K and G dwarfs to the M and F dwarfs studied in the previous section. This is because planets orbiting M dwarfs are expected to be in circular orbits and in this section we want to study the effects of planetary eccentric orbits. We selected the planet's semi-major axis so that the flux it would receive from the star at the apsis would be the same as the flux Earth receives from the Sun. Table \ref{tab:exomoons} shows the characteristics of the selected stars that we chose to represent each stellar type and the value of the semi-major axis obtained. The values quoted for Gliese 687 and HD 219134 were taken from \citet{pinotti2013most} and the values for HD 115383 were taken from \citet{da2012accurate}. The Sun's effective temperature was taken from \citet{prvsa2016nominal}. 

\begin{table}
	\centering
	\caption{Spectral type, mass, radius and effective temperature for the prototypical stars chosen to represent each one of the indicated types used in the simulation of the temperature forcing in hypothetical exomoons orbiting hypothetical gas giants in Section \ref{subsec:eccentricorbits}. The last column presents the value of our hypothetical planet's semi-major axis such that the stellar flux is the same as the flux Earth receives from the sun. }
	\label{tab:exomoons}
	\begin{tabular}{cccccc} 
		\hline
		Type & Star & $M_{\star}$ ($M_{\odot}$) & $R_{\star}$ ($R_{\odot}$) & $T_{eff} (K)$ & $a_p$ (AU) \\
		\hline
		M dwarf & Gliese 687 & 0.401 & 0.492   & 3095    & 0.141  \\
		K dwarf & HD 219134  & 0.850 & 0.684   & 5100    & 0.534  \\
		G dwarf & Sun        & 1     & 1       & 5772   & 1     \\
		F dwarf & HD 115383  & 1.22  & 1.305  & 6075    & 1.446  \\
		\hline
	\end{tabular}
\end{table}

For the F and G dwarfs, we applied our methodology for three values of planetary eccentricity, $e_P$ = 0, 0.3 and 0.6. For the M dwarf, we considered only a planet in circular orbit and for the K dwarf, we adopted eccentricities of 0 and 0.3. This choice was made because, specially for M dwarfs but also for K dwarfs, most orbits inside the HZ are expected to have been circularized and, also, the fact that the HZ is narrower and closer to the star makes planets with high eccentricities less interesting from an astrobiological point of view. The Hill radius for these cases ranges from 0.018 AU  to 0.133 AU, and the value of the radius for our hypothetical exomoons is within the stability zone in all cases studied \citep{domingos2006stable}. We studied the case of no atmosphere and albedo 0.25 and an Earth like atmosphere, described in Table \ref{tab:exoplanetsatm} as model 2, with $\chi = 3 \times 10^4$ mol/m$^2$ and albedo 0.3. In fact, although very thin atmospheres should be expected at those distances from the host star, we consider the possibility that the exomoons might have kept a somewhat thick atmosphere acquired at larger distances from the star, before planetary migration, if it is able to replenish its volatile inventory by continued volcanic activity, tidally induced.

\begin{figure}
	\includegraphics[width=\columnwidth]{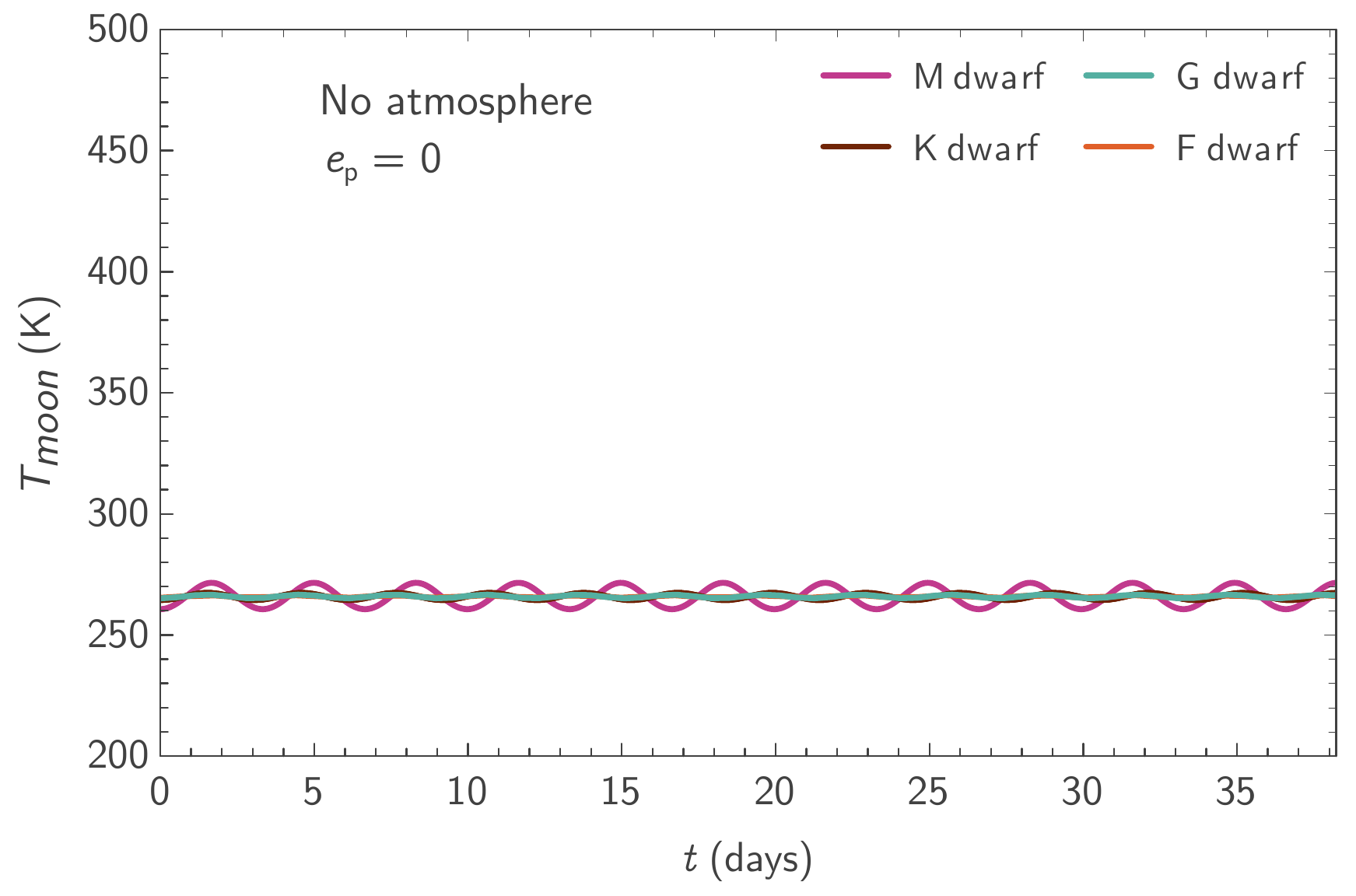}
	\includegraphics[width=\columnwidth]{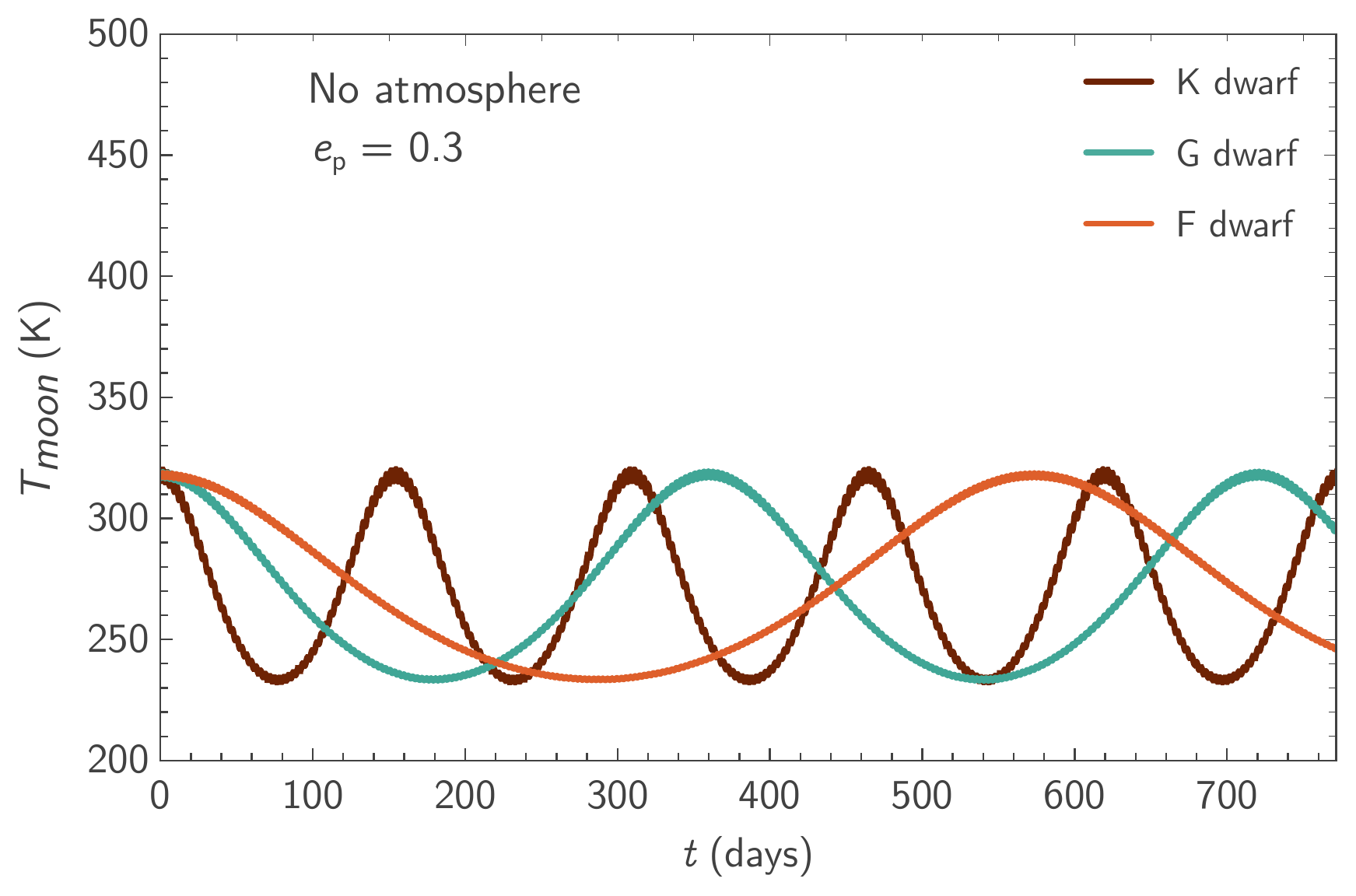}
	\includegraphics[width=\columnwidth]{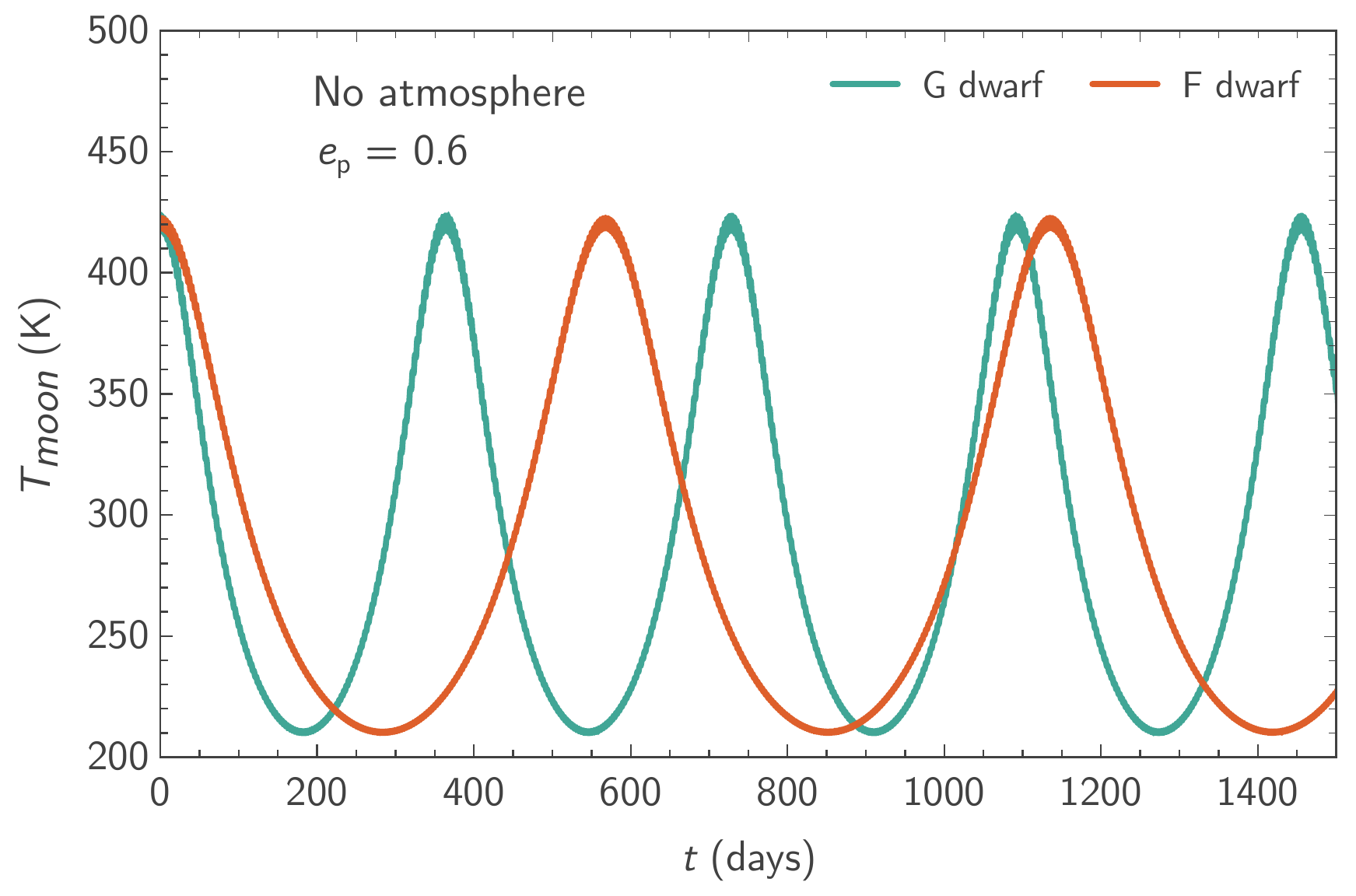}
	\caption{Brightness temperature profile for the hypothetical exomoons, with no atmosphere and albedo 0.25, studied in Section \ref{subsec:eccentricorbits}. Different colors represent moons in different planetary systems with an M dwarf, K dwarf, F dwarf and G dwarf. The upper panel shows the case for a planet in circular orbit ($e_p = 0$), the middle panel shows the same for a planetary eccentricity of $e_p$ = 0.3 and the lower panel shows the same for a planetary eccentricity of $e_p$ = 0.6.}    \label{fig:ExomoonsThin}
\end{figure}

\begin{figure}
	\includegraphics[width=\columnwidth]{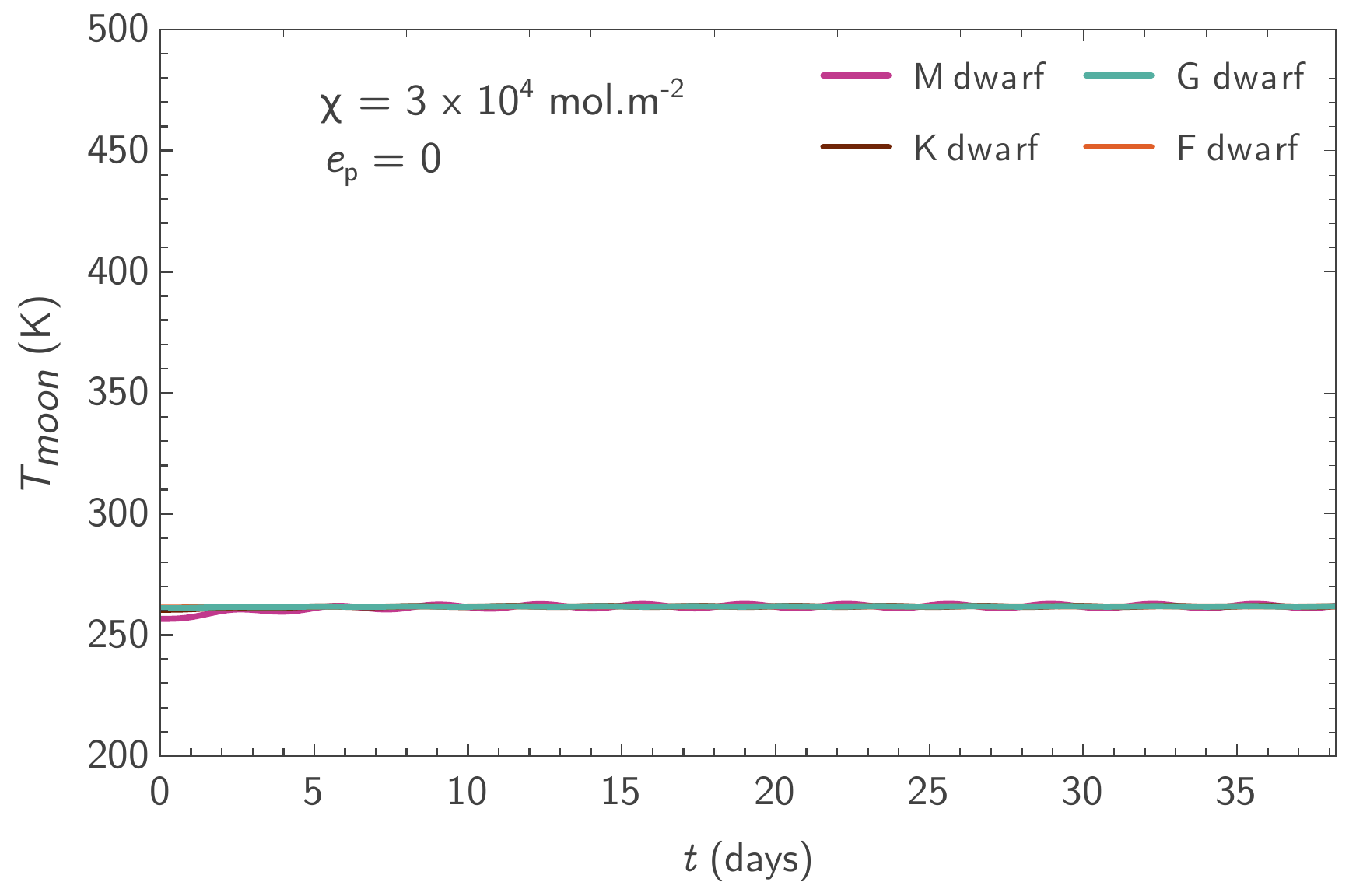}
	\includegraphics[width=\columnwidth]{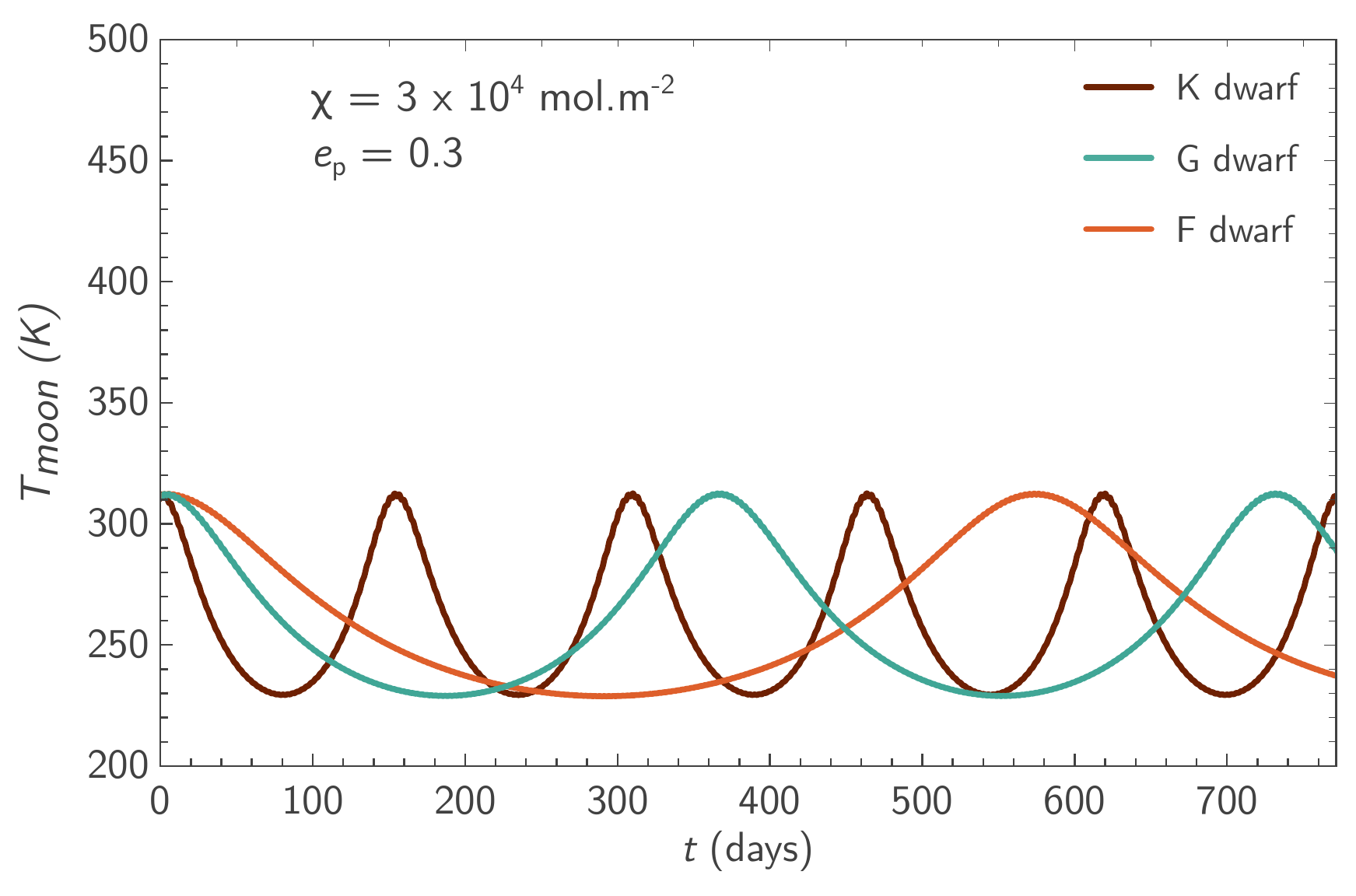}
	\includegraphics[width=\columnwidth]{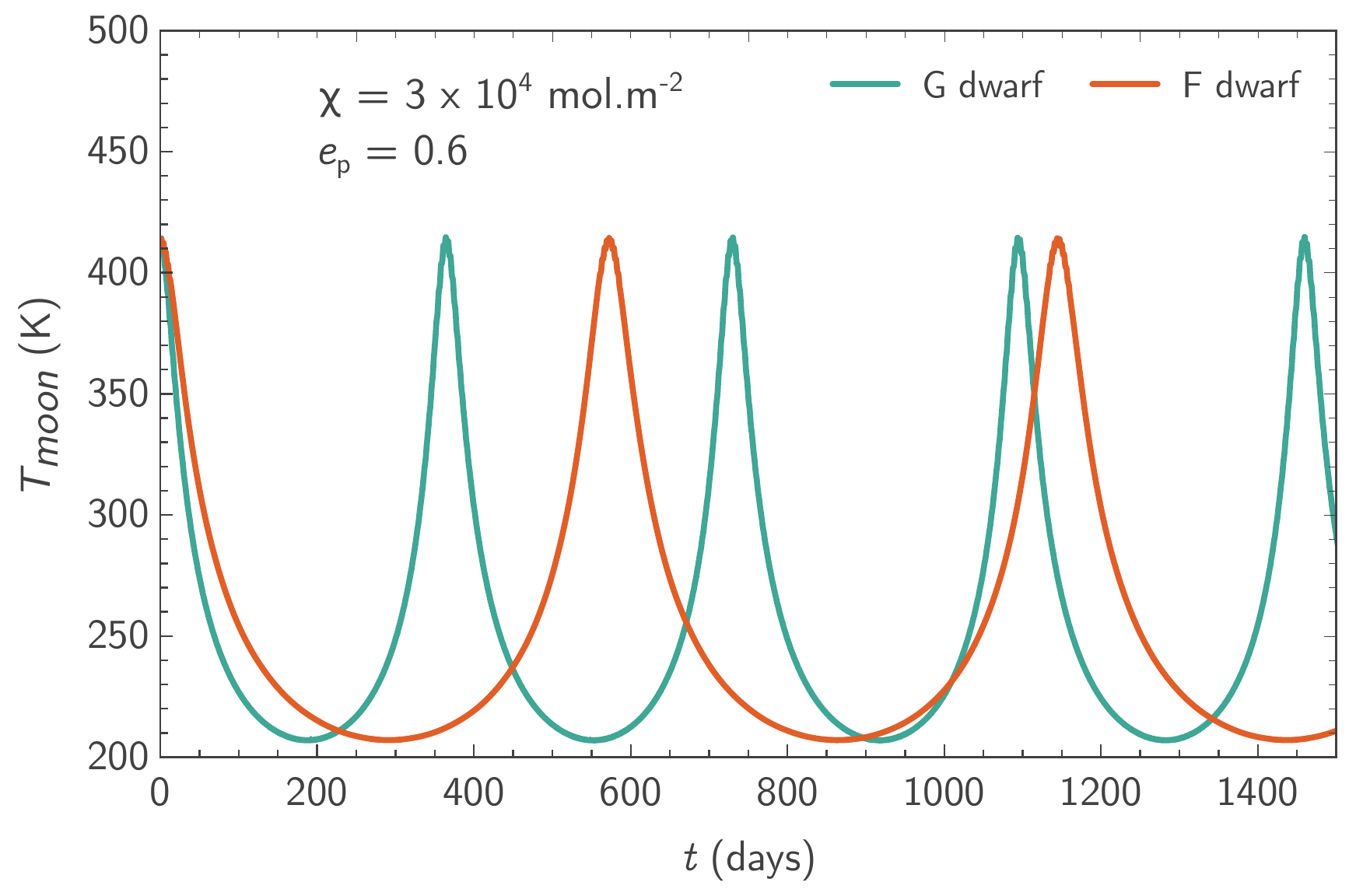}
	\caption{Same as Fig. \ref{fig:ExomoonsThin}, with an atmosphere ($\chi = 3 \times 10^4$ mol/m$^2$) and albedo 0.3.}    \label{fig:ExomoonsThick}
\end{figure}

The resulting temperatures can be seen in Fig. \ref{fig:ExomoonsThin}, for exomoons without atmosphere and albedo 0.25 and Fig. \ref{fig:ExomoonsThick} for the case of an atmosphere with $\chi = 3 \times 10^4$ mol/m$^2$ and albedo 0.3. In each plot, planetary eccentricity is fixed to the indicated value and different colors indicate the four different classes of stars chosen (pink for M dwarf, brown for K dwarf, orange for F dwarf and blue for G dwarf). We can notice that all the curves in a given panel vary around the same temperature value, since the planets' semi-major axis were chosen so that they would all receive the same flux from the star. The upper panel of Fig. \ref{fig:ExomoonsThin} shows the case of planetary circular orbits and moons with no atmosphere. In this case, there is no effect from the planet's orbit in the curves and the oscillations are due to the moon's orbit only. One can see that temperature variations are the highest for the case of M dwarf hosts (in pink) and this indeed should be expected, since the planet orbits much closer to the star than in the other cases. Notice that the orange curve in this panel is the same as the red curve in the upper panel of Fig. \ref{fig:CircularOrbits}. The other panels of Fig. \ref{fig:ExomoonsThin} show the cases of eccentric planetary orbits. As expected, temperature variations become larger for larger planetary eccentricities, ranging from $\sim 90$ K for the case of $e_P = 0.3$ to $\sim 210$ K for the more extreme situation of $e_P = 0.6$. Comparing the upper panels of Fig. \ref{fig:ExomoonsThin} and Fig. \ref{fig:ExomoonsThick}, we can once more clearly see how the presence of an atmosphere smooths temperature variations. For the case of $e_P = 0.3$, the resulting temperature variations are of $\sim 80$ K and, for $e_P = 0.6$, they are of $\sim 200$ K, 10 K lower than the corresponding cases with no atmosphere. 

\subsection{Hypothetical exomoons around real exoplanets} \label{subsec:realgiantplanets}

We finally tested our model for hypothetical moons around real giant exoplanets. The planets were selected to provide interesting case studies, spanning a range of stellar properties, varied planetary masses and also moderately large to large eccentricities, all of them orbiting within the HZ of the host stars. Tables \ref{tab:giant_exoplanets} and \ref{tab:giant_hosts} present some of their properties and properties of their host stars, respectively. All planets were detected by the radial velocity method and, therefore, we present the values of their minimum masses. The host stars include K, G and F-type objects, and the more massive stars were selected on the basis of showing evidence of still being at, or close to, the zero age main sequence. Thus we assume that the stellar HZ suffered little or no evolution, and the planets have been inside the HZ since their formation. Their range of eccentricities, $\sim$0.10-0.60, span essentially all cases of interest for our model. The minimum planetary masses are just under one Jupiter mass for the two K-stars in the sample, to at least three times as massive as Jupiter for one of the G-type stars.

\begin{table}
	\centering
	\caption{Identification (first column) and minimum mass, semi-major axis, orbital period and eccentricity (columns two to five, respectively) of the five giant exoplanets selected for examination of radiative forcing effects in hypothetical exomoons (Section \ref{subsec:realgiantplanets}).}
	\label{tab:giant_exoplanets}
	\begin{tabular}{ccccc} 
		\hline
		Name & $M_p$sin($i$) ($M_{J}$) & $a_p$ (AU) & $P$ (days) & e$_p$  \\
		\hline
		HD 128356 b & 0.89 & 0.87 & 298.2 & 0.57 \\
		HD 147513 b & 1.21 & 1.32 & 528.4 & 0.26 \\
		HD 165155 b & 2.89 & 1.13 & 434.5 & 0.20 \\
		HD 221287 b & 3.09  & 1.25 & 456.1 & 0.08 \\
		HD 45364 c & 0.66 & 0.90 & 342.9 & 0.10\\
		\hline
	\end{tabular}
\end{table}

\begin{table}
	\centering
	\caption{Host star properties for the exoplanets in Table \ref{tab:giant_hosts}: first and second columns give the spectral type and identification of the star; columns three, four and five, respectively, provide stellar mass, effective temperature and luminosity.}
	\label{tab:giant_hosts}
	\begin{tabular}{ccccc} 
		\hline
		Type & Name & $M_{\star}$ ($M_{\odot}$) & $T_{eff}$ (K) & $L_{\star}$ ($L_{\odot}$)  \\
		\hline
		K dwarf & HD 128356 & 0.65 & 4875 & 0.36   \\
		G dwarf & HD 147513 & 1.11 & 5883 & 0.98  \\
		G dwarf & HD 165155 & 1.02 & 5426 & 0.70  \\
		F dwarf & HD 221287 & 1.25  & 6304 & 1.66  \\
		K dwarf & HD 45364 & 0.82 & 5434 & 0.57  \\
		\hline
	\end{tabular}
\end{table}

\begin{figure}
	\includegraphics[width=\columnwidth]{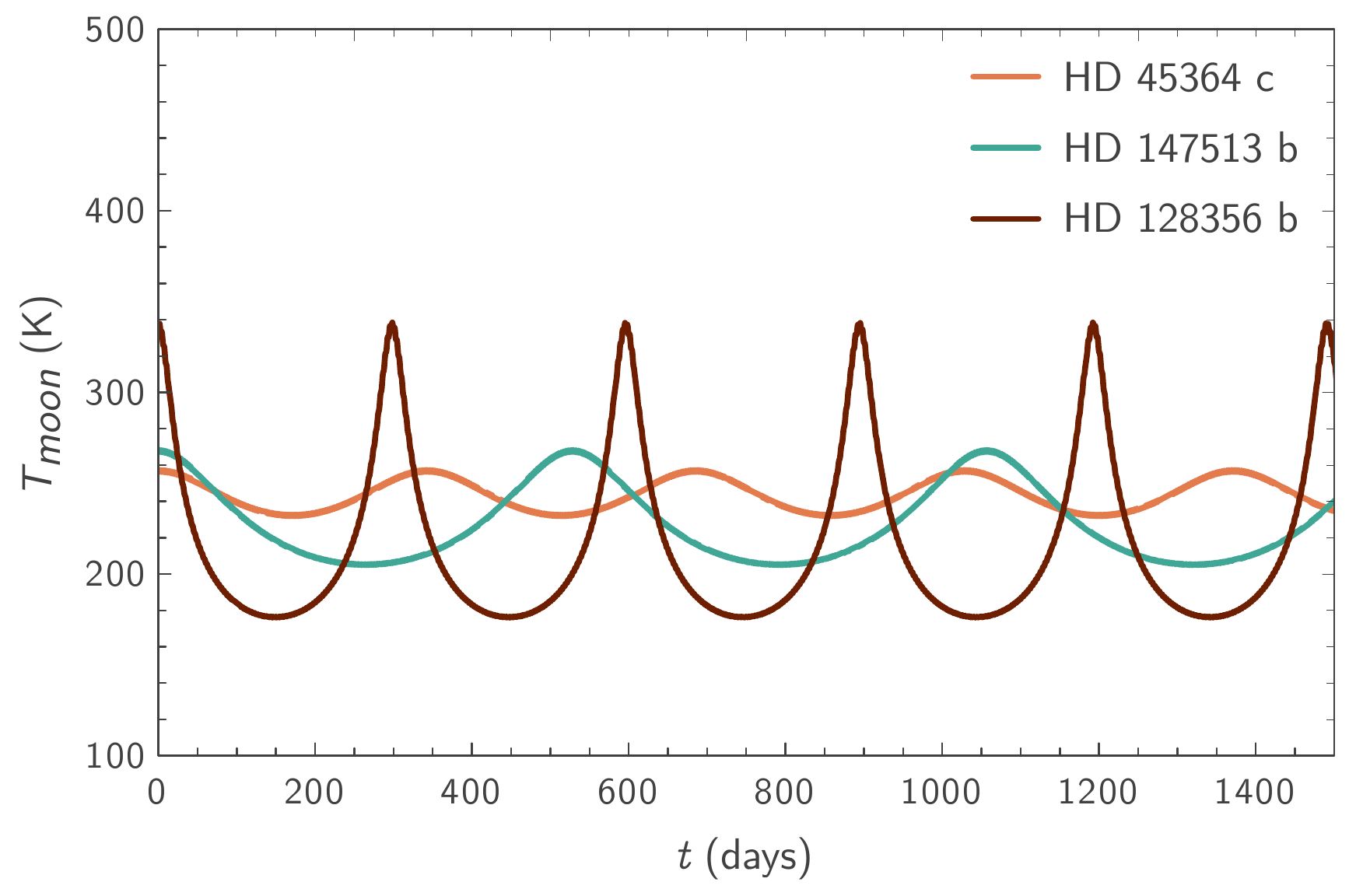}
	\caption{Brightness temperature profile for the hypothetical exomoons studied in Section \ref{subsec:realgiantplanets}, with Ganymede's mass, no atmosphere and albedo = 0.25, in circular orbits around exoplanets HD 128356 b, HD 147513 b and HD 45364 c.}    \label{fig:giants_ganymede}
\end{figure}

\begin{figure}
	\includegraphics[width=\columnwidth]{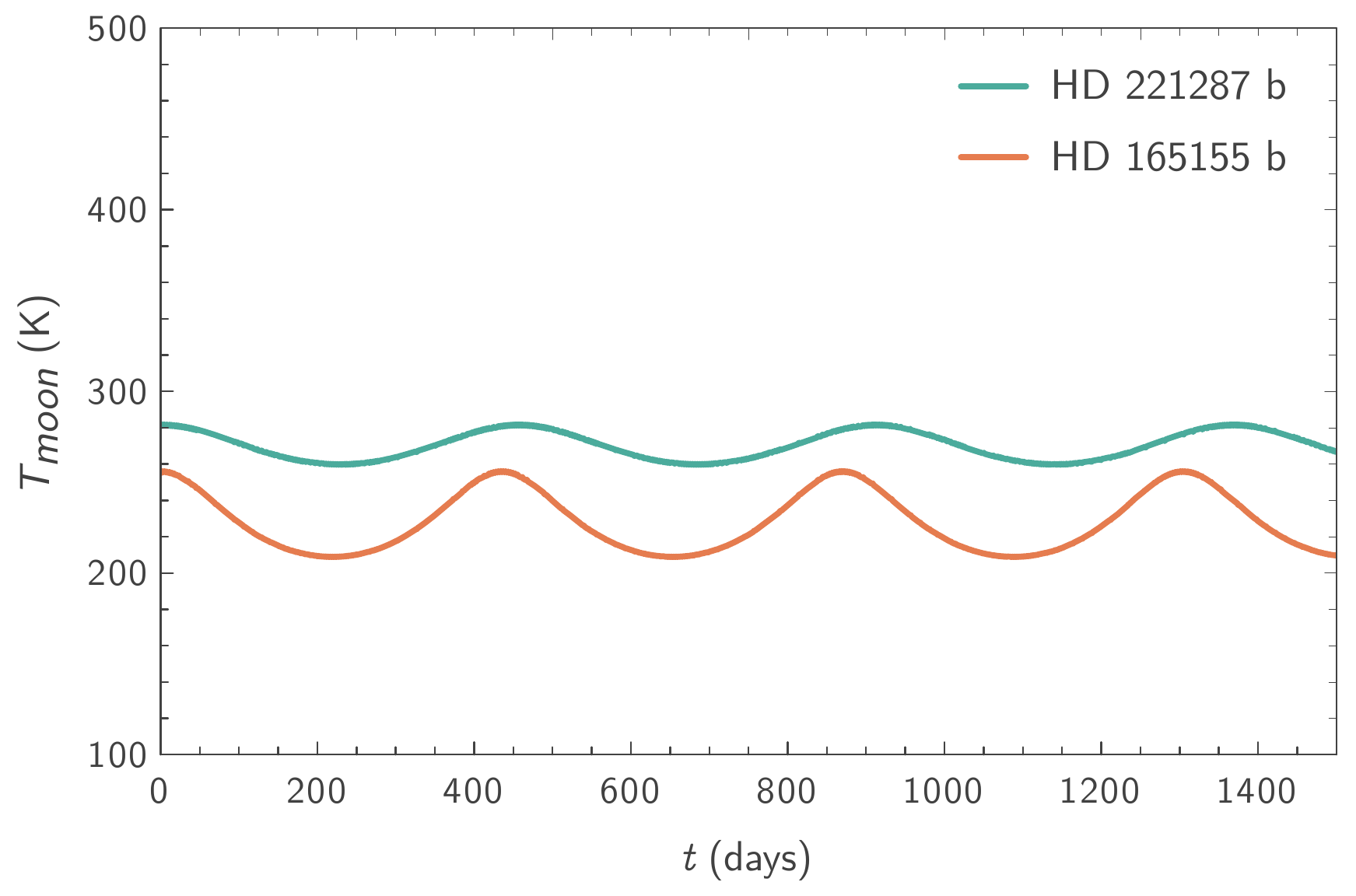}
	\caption{Brightness temperature profile for the hypothetical exomoons studied in Section \ref{subsec:realgiantplanets}, with Mars' mass, atmosphere with $\chi=3 \times 10^4$ mol/m$^2$ and albedo = 0.30, in circular orbits around exoplanets HD 165155 b and HD 221287 b. }    
	\label{fig:giants_mars}
\end{figure}

We separated the planets into two groups: the three less massive (HD 128356 b, HD 147513 b and HD 45364 c), with minimum masses less or around Jupiter's mass, and the two most massive (HD 165155 b and HD 221287 b), with minimum masses around three times Jupiter's mass. Following what was done in the previous Subsection, for the first group, we assumed a hypothetical exomoon with mass equal to Ganymede's mass, no atmosphere and albedo 0.25. For the second group, we studied a hypothetical exomoon with mass equal to Mars' mass, an Earth like atmosphere with $\chi = 3 \times 10^4$ mol/m$^2$ and albedo 0.30. In both cases, we calculated the exomoon's semi-major axis in order to produce 3-days orbital periods, keeping the eccentricity as null for all cases. The moons' orbital radii obtained vary from 0.0035 AU to 0.0058 AU, and we checked that, for each case, they are 4 to 6 times larger than the corresponding Roche limit and $\sim 20$ times smaller than the Hill radius. Figs. \ref{fig:giants_ganymede} and \ref{fig:giants_mars} show the temperature profiles obtained for each scenario.

The results show that temperature variations range from $\sim 20$ K (for the hypothetical exomoon orbiting HD 221287 b) to $\sim 170$ K (for the hypothetical exomoon orbiting HD 128356 b). This is readily understood by inspection of Table \ref{tab:giant_exoplanets}, where we see these planets have the lower and higher orbital eccentricities, respectively. Also, the fact that we model the exomoon orbiting HD 128356 b with no atmosphere makes higher temperature variations expected, with respect to the case of a similar moon with an atmosphere. The estimated temperature for the hypothetical exomoons around the other planets lie between 20 K and 170 K: $\sim 24$ K for HD 45364 c, $\sim 45$ K for HD 165155 b and $\sim 70$ K for HD 147513 b. Once again, we see that the order of increasing temperature variation follows the order of increasing planetary orbital eccentricity. Regarding the average temperature, the fact that the radiative, top of the atmosphere skin temperature for all five hypothetical exomoons vary around the freezing point of water $\sim 260-280$ K is due to the fact that all exoplanets are inside the HZ of their host star.

Uncertainties of the measured stellar and planetary parameters (with the exception of eccentricity) are mostly below $3\%$ and eccentricity measurements are the dominant source of error in the curves shown in Figs. \ref{fig:giants_ganymede} and \ref{fig:giants_mars}. The luminosity of HD 165155, $0.70 \pm 0.08\,L_{\odot}$, is an exception with a $\sim 10\%$ error bar \citep{2017MNRAS.466..443J}. Considering these uncertainties, the lower temperature variation of the hypothetical exomoon could reach $\sim 10$ K, for HD 221287 b's case,  and the higher $\sim 160$ K for HD 128356 b's case.

\section{Tidal dissipation in massive habitable exomoons} \label{sec:tidaldissipation}

The potential habitability of exomoons has long become an exciting topic in planetary science and astrobiology \citep{2022MNRAS.513.5290D,2020IJAsB..19..210L,2019ApJ...887..261M,2017A&A...601A..91D,2017MNRAS.472....8Z}, the chief reason being the expectation that massive exomoons are likely to exist around massive gas giants possibly up to one-third of an Earth mass \citep{2022MNRAS.513.5290D}. These large bodies could maintain liquid water at their surface for periods commensurate with the evolution of microbial life and, moreover, \cite{2021PASP..133i4401D} report that, for orbital periods compatible with the HZ of FGK stars, exomoons around giant planets are very likely to remain in stable orbits. Due to observational biases the first exomoons to be detected will be large, possibly more massive than Mars \citep{2013ApJ...776L..33H}.

Current evidence suggests that Mars itself was probably habitable for longer than 1 Gyr \citep[though perhaps only episodically,][]{2018NatGe..11..230R,2021NatGe..14..127W} and that volcanism was a relevant contributing factor \citep{2014NatGe...7...59R,2016AREPS..44..381W}. The maintenance of a geophysical internal engine is, probably, as relevant to providing long-lived habitable conditions on the surface of rocky exomoons as to supporting subsurface oceans or liquid habitats in icy ones. Subsurface biospheres may be longer lasting and perhaps even more benign, but such biospheres will be much harder to detect on exomoons than those based on surface conditions, which are the ones of interest to our approach.

Continued dissipation of tidal energy in large, rocky exomoons can be expected to maintain expressive volcanic activity, helping sustain a sizable atmosphere and maintain habitability, even in the absence of plate tectonics. Such activity might not be unlike the recently suggested protracted episodes of supervolcanism on Mars \citep{Whelley2021}, which may have lasted a few hundred Myr. Solar System data clearly indicates that there is a minimum required mass for rock bodies to hold on to a substantial fraction of their volatile content over long timescales, and that Mars-sized objects are below this threshold unless positioned very far from outer limit of the HZ, as testified by the case of Titan. The exomoon scenarios we invoke are severely affected as they assume low-mass objects inside the HZ and therefore closer to the host star than most large moons in our Solar System. Therefore, exogenous sources such as tidal dissipation enabling the maintenance of geologic activity and volcanism  are probably indispensable in allowing exomoons, within the HZ of FGKM stars, to keep substantial atmospheres over long timescales. Tidal dissipation has been found to be a relevant energy source for Mars-sized exomoons inside the HZ of such stars \citep{2017A&A...601A..91D,2017MNRAS.472....8Z}.

Towards evaluating some of the exomoon-exoplanet scenarios we investigated in Sec. \ref{sec:exomoons}, in terms of surface energy fluxes from tidal dissipation, compared to observed cases in our Solar System, we estimated the rates of energy dissipated per unit area at the surface of hypothetical exomoons, along one orbital period, from the formalism (their equations 4.77 and 4.197) of \cite{murray1999solar}:

\begin{equation}
\tilde{\mu}=\frac{19\,\mu}{2\,\rho\,g_p\,R_p}  , 
\end{equation}

\begin{equation}
\dot{E} = -\frac{63\,e^2\,n}{4\,\tilde{\mu}\,Q_p} \left( \frac{R_p}{a} \right)^5 \frac{G\,m_p^2}{a},
\end{equation}
where $\tilde{\mu}$ is the effective rigidity of the body, $\mu$ is the material rigidity, $\rho$, $g_{\rm p}$ and $R_{\rm p}$ are, respectively, the mean density, surface gravity and radius of the host planet, $\dot{E}$ is the dissipated power, {\it e} is the initial orbital eccentricity, $Q_{\rm p}$ is the thermal dissipation function, {\it a} is the orbital semi-major axis, $G$ is the gravitational constant and $m_{\rm p}$ is the planetary mass. The energy dissipated per area is denoted as $F_{\rm tidal}$ in Table \ref{tab:tidal_diss}.

We aim at providing single-body only, order of magnitude estimates in a preliminary exploration, so that the exomoon cases we examined in Sec. \ref{sec:exomoons} can be put into context. We do not consider details of the satellite architecture and interaction with companion exomoons. The values of $\mu$ were retrieved from \cite{murray1999solar} and correspond to $\mu_{\rm rocky}$ = 5.0 $\times$ 10$^{\rm 6}$ dyne.cm$^{\rm -2}$ for $\rho$ $\sim$ 3 g.cm$^{\rm 3}$. We reproduced the values of $\dot{E}$ given by \cite{murray1999solar} for the actual cases of Io, Europa and Ganymede to within a factor of two or better, differences probably being accounted for by different assumptions about $Q_{\rm p}$. Our estimates of $F_{\rm tidal}$ assume $Q_{\rm p}$ = 100, a value comparable to the one commonly cited for Mars ($Q_{\rm p}$ = 86) and are listed in Table \ref{tab:tidal_diss}, along with the corresponding values for Io, Europa and Ganymede.

\begin{table}
\centering
\caption{Tidal dissipation surface fluxes for representative Solar System cases and some scenarios investigated in the present work. First row are actual Solar System cases, second and third rows are hypothetical exomoon templates. First column is planetary mass in Jupiter masses; second column gives assumed mass for the hypothetical exomoons, equal to the masses of either Ganymede or Mars. Third and fourth columns, respectively, list initial orbital eccentricity assumed for the exomoons and tidal deformation energy fluxes (erg.cm$^{-2}$.s$^{-1}$).}
\label{tab:tidal_diss}
\begin{tabular}{cccc} \hline
$m_p$ ($M_J$) & Moon's template & $e$  & $F_{\rm tidal}$ \\ \hline
\multirow{3}{*}{Jupiter (actual)} & Io (actual)& 0.004 & 47.97 \\ \cline{2-4}
                                & Europa (actual)& 0.009 & 2.70 \\ \cline{2-4}
                                & Ganymede (actual) & 0.001 & 0.069\\ \hline
\multirow{2}{*}{1} & \multirow{2}{*}{Ganymede} & 0.001 &  3.33 \\ \cline{3-4}
                                               & & 0.094 &  29,148 \\ \cline{2-4}
                   & \multirow{2}{*}{Mars} & 0.001 & 0.66 \\ \cline{3-4}
                                           & & 0.094 &  5,760 \\ \hline
\multirow{2}{*}{3} & \multirow{2}{*}{Ganymede} & 0.001 & 30.5 \\ \cline{3-4}
                                               & & 0.094 &  263,723 \\ \cline{2-4}
                   & \multirow{2}{*}{Mars} & 0.001 & 5.97 \\ \cline{3-4}
                                           & & 0.094 & 52,050  \\ \hline
\end{tabular}
\end{table}

We consider hypothetical exomoons with masses either equal to those of Ganymede or Mars as templates, and located, as in Section \ref{sec:exomoons}, at distances corresponding to 3-day orbital periods (fast rotators) around gas giants with masses either equal to 1 $M_{\rm J}$ or 3 $M_{\rm J}$, and for initial eccentricities {\it e} = 0.001 and {\it e} = 0.094 (the same as Mars), totaling eight cases. Energy fluxes for the 1 $M_{\rm J}$ scenarios and low eccentricity cases are moderate and compare to the figures for the case of Europa or lower, but are three of four orders of magnitude higher for the higher initial eccentricity. High-eccentricity scenarios would converge to lower eccentricities by tidal relaxation, the details depending upon the structure of the satellite system the presence of Laplace resonances. Energy fluxes for the 3 $M_{\rm J}$ scenarios, however, even for the initially low eccentricity case, lie between the fluxes calculated for Europa and Io, and assume figures 10$^4$ higher for the higher initial eccentricity. Therefore, for exomoons approaching the mass of Mars and orbiting large gas giants, for which final eccentricities fall within the range observed for the Galilean satellites, significant tidal energy surface fluxes could be maintained over long timescales even allowing for tidal relaxation. An analysis of the long-term influence on volcanic activity of tidal surface fluxes higher than Europa's, for a rocky Mars-like body, has not been published to the best of our knowledge. For such and similar cases, important contributions to sustaining volcanic activity and a sizable atmosphere are likely, significantly improving the prospects for exomoon habitability.

\section{Conclusions} \label{sec:conclusion}

In this work we improve on the model of \cite{pinotti2013most}, originally developed for the study of radiative-forced temperature variations on exoplanets, extending the formalism (Sec.~\ref{sec:methods}) and exploring a wider range of cases for albedo, orbital eccentricity and atmospheric masses. We apply the model to estimate the brightness temperature modulation of both real and hypothetical rocky exoplanets and large exomoons, orbiting real and hypothetical host stars and host giant exoplanets inside the HZ. The results show that for specific cases an atmosphere may distort the exoplanet's or exomoon's profile of brightness temperature variations and serve as a measurable indication of the presence of an atmosphere. 

\begin{figure}
	\includegraphics[width=\columnwidth]{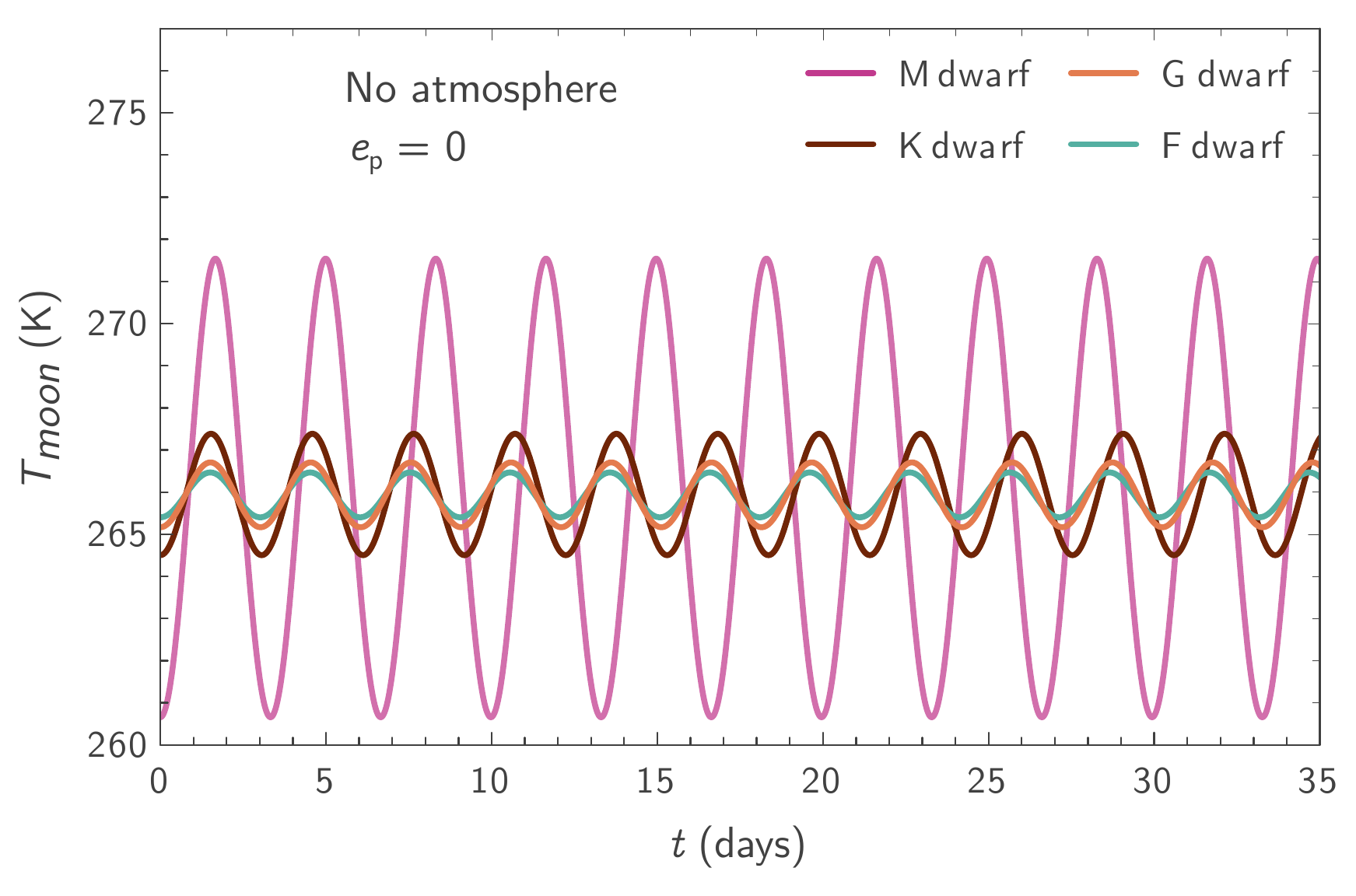}
	\caption{Zoomed-in version of the upper panel of Fig. \ref{fig:ExomoonsThin}, which shows the brightness temperature profile for the hypothetical exomoons with no atmosphere and albedo 0.25, orbiting a planet with circular orbit around four different types of stars (see Section  \ref{subsec:eccentricorbits}).}    
	\label{fig:zoomedin}
\end{figure}

We first analyze general model results for hypothetical Earth-like planets modeled on the Earth (Sec.~\ref{sec:earth}) and next apply it (Sec.~\ref{sec:exoplanets}) to two real exoplanets with orbital eccentricities {\it e} $\sim$ 0.1, for which the model yields temperature variations up to $\sim 24$ K along one full planetary orbit.

We next (Sec.~\ref{sec:exomoons}) employed Ganymede and Mars as templates for hypothetical exomoons and analyzed their epicyclic temperature brightness variations in circular orbit around planets with 1  and 3 Jupiter masses within the host star HZ. We also estimate (Sec.~\ref{sec:tidaldissipation}) tidal dissipation thermal fluxes, for various cases. We find substantial thermal fluxes at the exomoon surfaces for some cases. Such exomoons are potentially habitable and tidal dissipation provides a source of internal heating able to help maintain volcanism and thin atmospheres over extended timescales, possibly commensurate with those needed for biological evolution.

Our results show that the exomoons' temperature variations may reach values as high as $\sim 200$ K, for planetary orbital eccentricities {\it e} $\sim$ 0.6. Although such eccentricities are quite large, inspection of exoplanetary databases show that there are currently more than 100 exoplanets with measured eccentricities of 0.6 or larger, and 22 of which have masses larger than Jupiter's. HD 128356 b, which we analyzed in Section \ref{subsec:realgiantplanets} is an example of this case, with measured eccentricity of 0.57. The temperature brightness variation for an hypothetical exomoon in this case is $\sim 170$ K. 

For planetary eccentricities {\it e} $\sim$ 0.3, the exomoons' temperature variations may reach values of $\sim 90$ K. It is worth noting that more than 100 known exoplanets have orbital eccentricities larger than 0.3 and masses larger than Jupiter's ($\sim$ 80 of which have $e \leq 0.6$). Exoplanet HD 147513 b, studied in Section \ref{subsec:realgiantplanets} is a close example of this case, since it has a measured eccentricity {\it e} $=$ 0.26 and the temperature variation for the studied hypothetical exomoon around it reaches $\sim 70$ K. For the case of planetary circular orbits, analysed in Section \ref{sec:circular}, the maximum variation of temperature for the epicyclic motion of exomoons is $\sim 7-8$ K. The closest case to this situation is found to be HD 221287 b with $e=0.08$. A hypothetical exomoon with Mars' mass orbiting this planet would present temperature variations of $\sim 20$ K.

We analyzed cases of exomoons orbiting hypothetical planets (Secs.~\ref{sec:circular} and~\ref{subsec:eccentricorbits}) receiving from the host star the same flux that the Earth receives from the Sun. We compared resulting exomoon temperature variations for planets orbiting different types of stars, finding they are the highest for the case of M dwarfs, progressively decreasing for K, G and F stars in the case of moons with no atmosphere orbiting planets in circular orbits around their stars. This effect can be observed in Fig. \ref{fig:zoomedin}, which is a zoomed-in version of the upper panel of Fig. \ref{fig:ExomoonsThin}, where the moon has no atmosphere and both planet and moon describe circular orbits. Indeed, for the cases where the planetary orbits are circular, the same star-moon distance variation causes higher flux variations for close-in HZs. 

Such temperature variations do not significantly affect the star-moon contrast in these systems, which are around 10$^{-9}$-10$^{-8}$. In all cases studied by us, the temperature variations have too low a flux modulation effect to be detected by current instruments. Nevertheless, in the future, new instruments on ELT-class telescopes, like GMTNIRS/GMT \citep{2016SPIE.9908E..21J}, MODHIS/TMT \citep{2023AJ....165..113R} and HIRES/ELT \citep{2021Msngr.182...27M}, along with new space telescopes using IR nulling interferometry and upgrades to existing instruments at Keck, VLT and Subaru will provide the needed combination of high-sensitivity and high-angular resolution observations. Instruments such as ELT/METIS \citep{2018SPIE10702E..1UB}, combining coronagraphy for high contrast IR imaging and spectroscopy, will be able to perform direct characterization of exoplanets and possibly exomoons, finally bringing into reach the observations of temperature variations like those proposed by our work.

\section*{Acknowledgements}
The authors thank the anonymous referee for the enriching comments and suggestions. RGGF thanks the Brazilian Conselho Nacional de Desenvolvimento Científico e Tecnológico (CNPq) for the PIBIC undergraduate fellowship. GFPM acknowledges grant 474972/2009-7 from the Brazilian Conselho Nacional de Desenvolvimento Científico e Tecnológico (CNPq). MBF acknowledges financial support from he Brazilian Conselho Nacional de Desenvolvimento Científico e Tecnológico (CNPq) (grant number: 307711/2022-6). We thank Dr. Adrián Rodríguez for valuable discussions on tidal dissipation. The authors acknowledge the use of data from NASA’s planetary fact sheet webpage, produced by Dr. David R. Williams. This research has made use of the NASA Exoplanet Archive, which is operated by the California Institute of Technology, under contract with the National Aeronautics and Space Administration under the Exoplanet Exploration Program.

\section*{Data Availability}

The data underlying this article will be shared on reasonable request
to the corresponding author.

\bibliographystyle{mnras}
\bibliography{bibliografia}






\bsp	
\label{lastpage}

\end{document}